\documentclass[11pt]{article}
\usepackage{amsmath}
\usepackage{mathtools}
\usepackage{bbm}
\usepackage{color}
\usepackage{hyperref}
\usepackage[margin=1.0in]{geometry}
\usepackage{amssymb}
\usepackage{amsthm}
\usepackage{charter}
\usepackage{graphicx}
\usepackage{bm}
\newcommand{\vect}[1]{\boldsymbol{\mathbf{#1}}}

\DeclareMathOperator*{\argmin}{arg\,min}
\newcommand{\red}[1]{\textcolor{red}{#1}}
\DeclareMathOperator*{\der}{{\rm d}\!}
\DeclareMathOperator*{\E}{{\mathbb E}}

\title{Recent Advances in Modeling and Control of Epidemics using a Mean Field Approach} 

\author{Amal Roy, Chandramani Singh, and Y. Narahari \\Indian Institute of Science \\ Bangalore, India}
\begin{document}
\maketitle

\begin{abstract}
Modeling and control of epidemics such as the novel Corona virus have assumed paramount importance at a global level. A natural and powerful  dynamical modeling framework to use  in this context is a continuous time Markov decision process (CTMDP) that  encompasses  classical compartmental paradigms such as the Susceptible-Infected-Recovered (SIR) model. However, using a CTMDP based model suffers from the curse of dimensionality and moreover requires global state information.  
The challenges with CTMDP based models motivate the need for a more efficient approach and the {\em mean field approach\/}  offers an effective alternative.   The {\em mean field approach\/}  computes the collective behavior of a dynamical system comprising numerous interacting nodes (where nodes represent individuals in the population). This paper (a) presents an overview of the {\em mean field approach\/}  to epidemic modeling and control and (b) provides a state-of-the-art update on recent advances on this topic. 

When an epidemic strikes, it is important to contain and suppress the epidemic spread to minimize the loss of lives and the suffering as well as  lower the burden on the public health care system. Non-pharmaceutical interventions 
are simple yet  effective measures to limit the spread of an ongoing epidemic. However, it is observed that in spite of the threat posed by an epidemic, individuals tend to go by their freewill rather than adhering to universally accepted best practices.  This could dangerously push the population level  behavior towards alarming consequences.  
 Motivated by this, our discussion in this paper proceeds along two specific threads. The first thread assumes that the individual nodes faithfully follow a socially optimal control policy prescribed by a regulatory authority. The second thread allows the individual nodes to exhibit independent, strategic behavior. In this case, the strategic interaction is modeled as a mean field game and the control is based on the associated mean field Nash equilibria.  
 In this paper, we start with a discussion of modeling of epidemics using an extended compartmental model - SIVR  (Suceptible-Infected-Vaccinated-Recovered) and provide an illustrative example.
 We next provide a review of relevant literature, using a mean field approach, on optimal control of epidemics, dealing with how a regulatory authority may optimally contain epidemic spread in a population. Following this, we provide an update on the literature on the use of the mean field game based approach in the study of epidemic spread and control. We conclude the paper with relevant future research directions. 

\end{abstract}

\section{Introduction}
Throughout history, epidemics have had major effects, often catastrophic, on the lives and lifestyle of the global population. The {\em Bubonic Plague\/} also called {\em Black Death\/} ravaged Asia and Europe in several waves during the fourteenth century, and is estimated to have caused the death of as much as one third of the population of Europe between 1346 and 1350. {\em Spanish Flu\/}, also known as the {\em Great Influenza\/} epidemic, was an exceptionally deadly global influenza pandemic which started in 1918.  Two years later, nearly a third of the global population, or an estimated 500 million people, had been infected in four successive waves. Estimates of deaths range from 17 million to 50 million, and possibly as high as 100 million, making it one of the deadliest pandemics in human history. 

More recently, the entire world has been devastated by  multiple waves of a novel Corona virus and its variants since December 2019. Unsurprisingly, this has led to research and innovation efforts of an unprecedented scale towards prediction, mitigation, and management of the pandemic.  Epidemiology has now moved to the centre-stage of research and policy making in public health.  \\[1mm]

\noindent
{\bf Motivation}. When an epidemic or pandemic such as Corona strikes, it is important to contain and suppress the spread of the disease to minimize the loss of life and the suffering and also to lower the burden on the health care system. Non-pharmaceutical interventions such as reduction of social interactions, masking, social distancing, hand washing, hand-sanitizing, and disinfecting surfaces are simple yet very effective measures to limit the spread of an ongoing epidemic. However, it is observed that in spite of the threat posed by the pandemic individuals tend to go by their freewill rather than adhering to  best practices such as stated above. There are numerous instances, throughout the world, where individuals have refused to comply with Covid appropriate behaviour \cite{AURELL21a}. This could dangerously push the  dynamics of the entire population towards alarming or even catastrophic consequences.  This lack of responsible behavior has forced regulatory authorities to seek suitable  measures and incentives to improve compliance to best practices. In order for these measures to be computed scientifically, we need an appropriate modeling and control framework.  \\[1mm]

\noindent
{\bf Mean Field Approach}. Set in the above backdrop, modeling and control of epidemics have assumed paramount importance. A natural and powerful model that has emerged in this context is a continuous time Markov decision process (CTMDP) which encompasses the classical compartmental paradigms such as the SIR (Susceptible-Infected-Recovered) model.  The use of  a CTMDP based model, however,  poses certain technical and computational challenges. First, the state space of any CTMDP based model grows exponentially with the population size and the computation of the optimal policy using dynamic programming principles become intractable even for moderate population sizes \cite{sutton2018reinforcement}. Second, the controller needs to know the global population state to execute the policy. These two challenges motivate the need for a more efficient approach and the {\em mean field approach\/}  has emerged as an effective alternative \cite{GAST12}. The mean field approach computes the collective behavior of a dynamical system comprising numerous interacting nodes (individuals in the population). 
Our objective in this paper is to provide a bird's eye view  of and a state-of-the-art update on some recent advances in the use of the mean field approach for epidemic modeling and control. \\[1mm]

\noindent
{\bf Structure of the Paper}. Our discussion in this paper proceeds along two threads. The first thread assumes that the individual nodes faithfully follow a socially optimal control policy prescribed by a regulatory authority. Here we follow a mean field approach to derive optimal control. The second thread allows the nodes to exhibit individualistic, strategic behavior. In this case, a mean field game is formulated and the individuals' controls are governed by the mean field Nash equilibrium.  This paper is structured as follows. 


\begin{itemize}
    \item {\bf Section 2}: This  section is devoted to a description of the model of spread of epidemics that we will be using in the rest of the paper. First, we present an extension of the classical SIR model (Susceptible-Infected-Recovered model) taking into account vaccinations. We call this the SIVR (Susceptible-Infected-Vaccinated-Recovered) model. We describe the costs incurred by each node: lockdown cost, infection cost, and  vaccination cost. We next describe the evolution of the population, which turns out to be a time-inhomogeneous continuous time Markov chain. We formulate the optimal control problem which seeks to minimize the expected average cost for an individual over a finite time horizon. We next  present a stochastic game model when the individual nodes are strategic.
    \item {\bf Section 3}: This discusses mean field modeling of epidemics. The spread of epidemics can be modeled as the mean field limit of a sequence of dynamical processes. We consider a general dynamical system and recall key results in mean field analysis. We present the standard solution to the mean field control problem. We then consider the strategic case and present the mean field game model and a solution for the same.
    \item {\bf Section 4}: Here, we present an illustrative example. To keep things simple, we consider an SIR model rather than an SIVR model. For this example, we illustrate mean field control for the non-strategic case followed by a mean field game model and control for the strategic case.
    \item  {\bf Section 5}: This provides a state-of-the-art update of mean field optimal control of epidemics when a centralized regulatory authority prescribes a control policy and the individual nodes faithfully follow the policy. There is rich literature on applying optimal control methods to compartmental models of epidemics. We categorize the literature into 
(1) Non-pharmaceutical interventions (2) Vaccination strategies. 

\item {\bf Section 6}: This section is devoted to a  state-of-the-art update of literature on the use of the mean field game approach in  the study of epidemic spread and control. We have categorized the relevant literature into the following groups: (1) Non-pharmaceutical interventions (2) Vaccination strategies (3) Control and policy design.
\item {\bf Section 7}: We conclude the paper  by providing several directions for future research in this section. 
\end{itemize}



\section{Modeling of Epidemics} 
\label{sec2}
Modeling of spread of epidemics is a vast topic and there are numerous papers which present a variety of models.
We wish to point to the paper by Guan, Wei, Zhao, and Chen \cite{GUAN2020} which provides a review of literature on modeling techniques and dynamic models in the context of the COVID-19 pandemic. The authors there discuss how three aspects, (a) epidemiological parameter estimation, (b) trend prediction, and (c) control measure evaluation, which are addressed in the literature. They conclude that dynamic models which are extensions of basic SIR (susceptible-Infected-Recovered) and SEIR (Susceptible-Exposed-Infected-Recovered) models provide useful insights into these three aspects. Their conclusion is that dynamic models are useful for exploring possibilities of interventions but can fail at making strong predictions about long-term disease dynamics. 

In this section, we present the modeling framework that we will be using in the rest of the paper. We first describe an extension of the classical SIR   (Susceptible-Infected-Recovered model) compartmental model taking into account vaccinations. We call this the SIVR (Susceptible-Infected-Vaccinated-Recovered) model. We find this compartmental model congenial for illustrating the computation of optimal control and mean field equilibrium. We describe the costs incurred by each node: lockdown cost, infection cost, and  vaccination cost. We next describe the evolution of the population, which turns out to be a time-inhomogeneous continuous time Markov chain. We formulate the optimal control problem which seeks to minimize the expected average cost for an individual over a finite time horizon. We then present a mean field game model where the individual nodes are strategic. 
Important symbols used in this paper are listed for ready reference in Table 1.
\begin{table*}[t]
    \centering
    \begin{tabular}{ ||c|c| } 
 \hline \hline
 $N$ & Number of nodes (individuals or agents) in the population  \\ \hline
 $\mathcal{S}$ & State space of a Markov chain \\ \hline
 $K$ & Cardinality of a discrete state space $\mathcal{S}$ \\ \hline
  $T$ & Time horizon  \\ \hline
 $M_i$ & Number of nodes in state $i$  \\ \hline
 $Y_i$ & State of agent $i$. $ Y_i \in \mathcal{S}$  \\ \hline
 $X_i$ & $X_i:=M_i/N$, fraction of nodes in state $i$  \\ \hline
 $p_i$ & Probability of an agent being in state $i$  \\ \hline
   $\Delta^{N}_K$ & $\{ \frac{x}{N} | x \in \mathbb{Z}^{K}_{+},  \sum_{i=1}^K x_i = N \}$ \\ \hline
 $\Delta_K $ & $\{ x \in \mathbb{R}^K_{+} | \sum_{i=1}^K x_i =1 \} $ \\ \hline
  $X^{N}$ & State of population of $N$ agents $:=(X_1,\ldots,X_k) \in \Delta^N_K$  \\ \hline
 $x(t)$ & State of the population in mean field $:=(x_1,\ldots,x_k) \in \Delta_K$  \\ \hline
 $Q$ & $\in \mathbb{R}^{K \times K}$ Transition rate matrix. $Q_{ij}$ is the transition rate from state $i$ to $j$ \\  \hline
 $c_L (\cdot)$ & Lockdown cost per unit time per individual \\ \hline
 $c_I (\cdot)$ & Infection cost per unit time per individual \\ \hline
 $c_V$ & Vaccination cost per individual \\ \hline
 $u(\cdot)$ & Control variable as a function of time. $u(t) \in U \subset \mathbb{R}^m$ \\ \hline
 $g(\cdot,\cdot)$ & Running cost function for optimal control problem. $g: \mathbb{R}^K \times U \rightarrow \mathbb{R}$ \\ \hline
 $h(\cdot)$ & Terminal cost function for optimal control problem. $h: \mathbb{R}^K \rightarrow \mathbb{R}$ \\ \hline
 $\pi_t$ & Markov decision rule $\pi_t : \Delta^N_K \rightarrow U$ \\ \hline
 $\pi$ & $\pi := (\pi_t, t \in [0,T]) $ is the policy  \\ \hline
 $J_{\pi}(x)$ & Cost for the population, given the initial state $x$ and policy $\pi$\\ \hline
 $J_{\bar{\pi}}(i,x,\pi)$ & Cost for an agent starting in state $i$, using policy $\bar{\pi}$, with the population \\ & starting in state $x$ and using policy $\pi$\\
 \hline \hline
\end{tabular}
    \caption{List of symbols and notation.}
    \label{tab:list}
\end{table*}




\subsection{Modeling an Individual Node: SIVR Model}

\begin{figure}[h]
    \centering
    \includegraphics[width=0.5\textwidth]{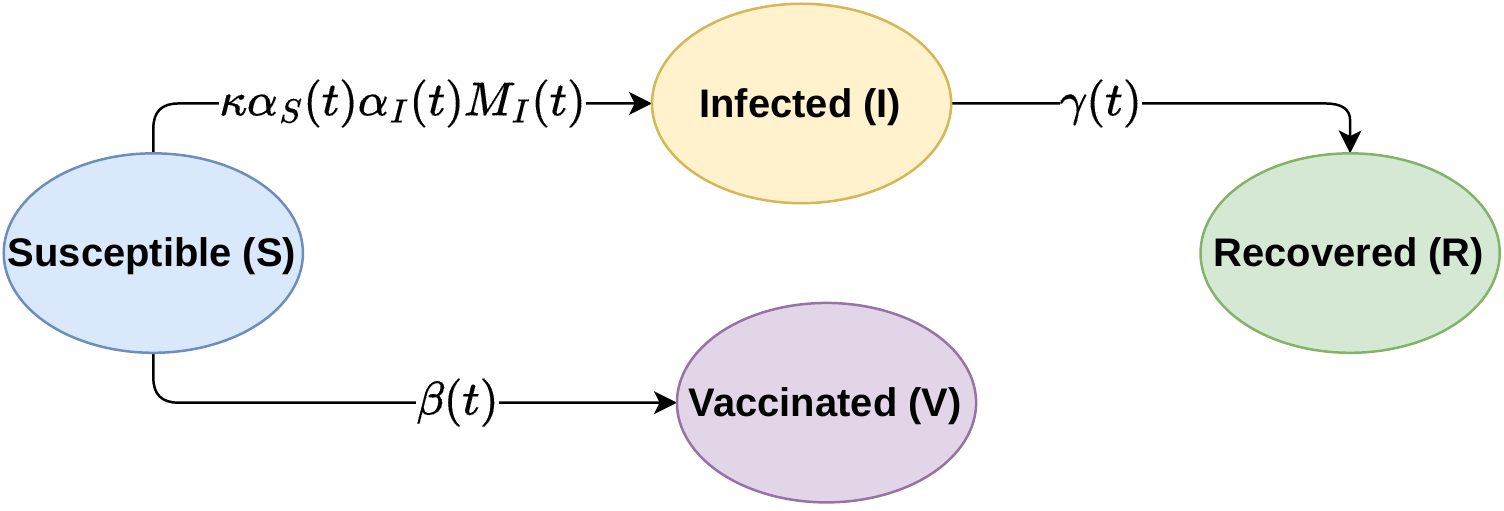}
    \caption{State transition diagram for an individual node. Values in edges are the transition rates between states.}
    \label{fig:transition_individual}
\end{figure}

Let us consider a population of $N$ nodes. Each node can be in one the four states, {\it susceptible}~($S$), {\it infected}~($I$), {\it vaccinated}~($V$) or {\it recovered}~($R$). The state transition diagram for an individual node is shown in Figure \ref{fig:transition_individual} (a more detailed description of the figure appears subsequently). The nodes' states evolve with time due to their interactions with other nodes, external interventions or because of their own actions. For instance, a susceptible node becomes infected on meeting an infected node or can choose to become vaccinated. On the other hand, an infected node may recover over time either on its own or through medical treatment. We assume that recovered nodes will not be infected in future. 

\begin{figure}[h]
    \centering
    \includegraphics[width=0.5\textwidth]{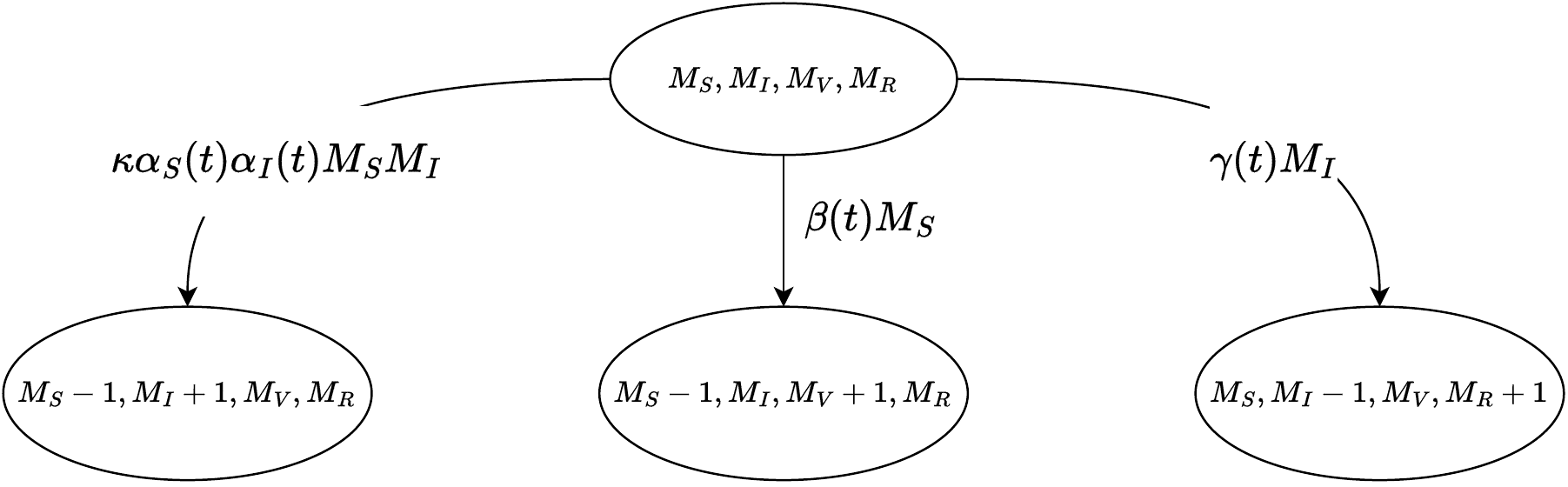}
    \caption{State transition diagram of the continuous time Markov chain $M(t), t \geq 0$ at $M(t) = (M_S,M_I,M_V,M_R)$.}
    \label{fig:transition_system}
\end{figure}

We assume that the number of unrestricted meetings between any pair of nodes constitute a  Poisson process of rate $\kappa$ and that these processes are independent across the node pairs. Moreover, the number of meetings between  any pair of susceptible-infected nodes constitutes a Poisson process of rate $\kappa\alpha_S(t)\alpha_I(t)$ where $\alpha(t) \coloneqq (\alpha_S(t), \alpha_I(t)) \in [\alpha_{\min}, 1] \times [\alpha_{\min}, 1] \eqqcolon A$ quantify the efforts of susceptible and infected individuals to contact other individuals. We also assume that the susceptible nodes' vaccination processes are independent Poisson processes with rate $\beta(t) \in B$ and infected nodes' recovery times are exponentially distributed with parameter $\gamma(t) \in C$. We refer to $u(t) \coloneqq (\alpha(t),\beta(t),\gamma(t)) \in U \coloneqq A \times B \times C$ as the action at time $t$, and the function $u : [0,T] \to U$ as the action function. 

\subsection{Evolution of the Population}
\label{sec2.2}
Let $Y_n(t)$ denote the state of the $n^{th}$ node at time $t$ and $Y(t)$ denote the state of the entire population; $Y(t) \coloneqq (Y_1(t), Y_2(t),\cdots, Y_N(t))$. Let $M(t) \coloneqq (M_S(t),M_I(t),M_V(t),M_R(t))$ denote the numbers of nodes in states $S,I,V$ and $R$ at time $t$, respectively;
\[
M_i(t) = \sum_{n=1}^N \mathbbm{1}_{\{Y_n(t)=i\}} \ \text{for} \ i \in \{S,I,V,R\}.
\]
From the above discussion, $Y(t), t\geq 0$ is a time inhomogeneous continuous time Markov chain~(CTMC). For the $n^{th}$ node, its  state evolution at any time $t$ depends on the joint population state $Y(t)$ only through $Y_n(t)$ and $M(t)$.  Figure~\ref{fig:transition_individual} shows the associated state transition rate diagram. 
Moreover, $M(t), t\geq 0$ is also a CTMC with transition rate diagram as in Figure~\ref{fig:transition_system}.
Observe that $M_S(t)+M_I(t)+M_V(t) + M_R(t) = N$ for all $t$, and hence, it suffices to specify three numbers, $M_S(t),M_I(t)$ and $M_V(t)$.

Further, let $X(t) \coloneqq (X_S(t),X_I(t),X_V(t),X_R(t))$ denote the fractions of nodes in states $S,I,V$ and $R$ at time $t$; 
\[
X(t) = \left(\frac{M_S(t)}{N},\frac{M_I(t)}{N},\frac{M_V(t)}{N},\frac{M_R(t)}{N}\right) \in \Delta^N_4
\]
where $\Delta^N_K \coloneqq \{\frac{m}{N}: m \in \mathbb{Z}^K_+,  \sum_{i=1}^K m_i =N\}$. Clearly, $X(t), t\geq 0$ is also a CTMC.
Both $M(t), t\geq 0$ and $X(t), t\geq 0$ offer equivalent characterisations of the same dynamical system. From the state transition diagram in Figure \ref{fig:transition_system}, we can also write the  expected conditional drift rates of $X(t)$. These are as follows:
\begin{subequations} \label{eq:state_drift}
\begin{align}
\frac{\der \mathbb{E}[X_S(t) |X(t)]}{\der t} & =   \kappa\alpha_S(t)\alpha_I(t) (N X_S(t))(N X_I(t)) \left(\frac{-1}{N}\right)
\nonumber \\ & \quad + \beta(t) N X_S(t) \left(\frac{-1}{N}\right) \nonumber \\
& =  -N\kappa\alpha_S(t)\alpha_I(t)X_S(t) X_I(t) - \beta(t) X_S(t).  \label{drift_S}    \\
\frac{\der \mathbb{E}[X_I(t) |X(t)]}{\der t} & =   N\kappa\alpha_S(t)\alpha_I(t) X_S(t) X_I(t) - \gamma(t) X_I(t)  \label{drift_I}. \\
\frac{\der \mathbb{E}[X_V(t) |X(t)]}{\der t} & =   \beta(t) X_S(t)
\label{drift_V}. 
\end{align}
\end{subequations}
We prefer to 
work with $X(t), t\geq 0$ for reasons that will be clear in the next section. 

\subsubsection{Costs Incurred by the Nodes}
We characterise the costs incurred by the nodes during the course of an epidemic. 
An individual can incur the following costs, according to the action at time $t$, $u(t)=(\alpha(t),\beta(t),\gamma(t)) \in A \times B \times C$.
\begin{enumerate}
    \item {\em Lockdown cost:} This quantifies financial costs or discomfort caused to a susceptible or infected individual due to social distancing and lockdowns. These are reflected in susceptible and infected nodes' controls of meeting rates $\alpha_S(t)$ and $\alpha_I(t)$, respectively. We define a function $c_L: [\alpha_{\min},1] \rightarrow \mathbb{R}$ where $c_L(\alpha_S(t))$ and $c_L(\alpha_I(t))$ represent the lockdown costs per unit time of susceptible and infected nodes, respectively.
    \item {\em Infection cost:} This captures medical expenses, financial losses, etc., incurred by an infected individual. In general, the infection cost per unit time and the average recovery time $1/\gamma(t)$ are correlated. For instance, better health care incurs more cost per unit time but facilitates quicker recovery. To capture this correlation, we define function $c_I: C \rightarrow \mathbb{R}$ and $c_I(\gamma(t))$ denotes the infection cost per unit time.
    \item {\em Vaccination cost}: This represents the per individual cost of vaccination and treatment of side effects of vaccination, if any. We assume the vaccination cost to be fixed.  Let $c_V$ represent the vaccination cost per individual.
\end{enumerate}
Now we compute the average total cost, averaged over all the individuals. We express this cost as a function of initial population distribution $X(0) = x$ and actions $u(t) \coloneqq (\alpha(t), \beta(t), \gamma(t)), t \geq 0$. 

\begin{align}
&\frac{1}{N}\sum_{n =1}^{N}\E\left[\right.\int_0^T \left(c_L (\alpha_S(t)) \mathbbm{1}_{Y_I (t) S} +  (c_L(\alpha_I(t))+c_I(\gamma(t))) \mathbbm{1}_{Y_I (t) = I}\right) \der t \nonumber \\
& \qquad  \qquad + c_V \mathbbm{1}_{Y_I (T) = V}\ \vert \ X(0) = x\left.\right] \nonumber \\
& = \E\left[\right.\int_0^T \left(c_L (\alpha_S(t))X_S(t) + (c_L(\alpha_I(t)) +c_I(\gamma(t))) X_I(t)\right) \der t \nonumber \\
& \qquad  \qquad  +  c_V X_V(T)\ \vert \ X(0) = x\left.\right]
\label{eq:exp_cost}
\end{align}

\subsubsection{The Optimal Control Problem}
Observe that population evolution as well as the expected average cost over $[0, T]$ depend on actions $u(t), t \in [0,T]$. The optimal control problem seeks to minimize this cost via appropriate choice of actions. 
\paragraph{Decision rule:} A decision rule at time $t$, $\pi_t$, is function that takes the states $X(s), s \in [0,T]$ as input and output the action or decision $u(t)$. A history dependent decision rule at time $t$ relies only on $X(s), s \in [0,t]$ for its output. Further, the output of a Markov decision rule at time $t$ depends only on the instantaneous state $X(t)$. More precisely, a Markov decision rule $\pi_t$ is a mapping $\pi_t: \Delta^N_4 \to U$.

\paragraph{Policy:} A policy is an ordered set of decision rules for all $t \in [0,T]$:
\begin{equation} \label{eq:policy}
    \pi \coloneqq (\pi_t, t \in [0,T])
\end{equation}
A policy is called history dependent~(respectively, Markov) policy if it consists of history dependent~(respectively, Markov) decision rules.

We have at our disposal a finite horizon continuous time Markov decision process~(CTMDP) with finite state space and compact action spaces. Furthermore, the cost rates are bounded. Hence, following \cite{GAST12}, there exists a Markov policy that is optimal within the class of all deterministic history-dependent policies. Further, observe that a Markov policy $\pi$ together with the initial state $X(0)$ characterize the whole trajectory $X(t), t \in [0,T]$. So, the cost in \eqref{eq:exp_cost} also is a function of $\pi$ and $X(0) = x$, we use $J_{\pi}(x)$ to refer to this cost,
\begin{align}
    \lefteqn{J_{\pi}(x)} \nonumber \\
& \coloneqq \E\left[\right.\int_0^T \left(\right.c_L (\alpha_S(t))X_S(t) + (c_L(\alpha_I(t)) +c_I(\gamma(t))) X_I(t)\left.\right) \der t \nonumber \\
& \qquad+  c_V X_V(T) \ \vert \ X(0) = x\left.\right]
    \label{eq:opt_cost_j}
\end{align}
where $(\alpha(t),\beta(t),\gamma(t)) = \pi_t(X(t))$. The cost-to-go from a starting state $x \in \Delta^N_4$ is given by \eqref{eq:value_pol_opt1} and the optimal control is given by \eqref{eq:value_pol_opt2}. The following pair of equations solves the Markov control problem.
\begin{subequations} \label{eq:value_pol_opt}
\begin{align}
    J(x) &= \min_{\pi \in \Pi} J_\pi(x) \label{eq:value_pol_opt1} \\
\pi^\ast & \in \argmin_{\pi \in \Pi} J_\pi(x) . \label{eq:value_pol_opt2}
\end{align}
 \end{subequations}
Here $\Pi$ is the set of all Markov policies. However, this solution has two drawbacks.
\begin{enumerate}
    \item The state space of the CTMDP grows exponentially with the population size. Consequently, the algorithm to obtain the optimal policy has a prohibitive complexity even for moderate population sizes.  
    \item The controller needs to know the global population state to execute the policy.  
\end{enumerate}
Both these problems are addressed with a mean-field approach. We work with the mean-field limit of the CTMDP as described in Section \ref{sec3}.

\subsection{Evolution of an Individual Node}
 We now focus on a particular node, say the $n$th node. Let $p(t) \coloneqq (p_S(t), p_I(t),p_V(t), p_R(t))$ denote the conditional probabilities of this node being in states $S$, $I$, $V$ and $R$, respectively, at time $t$, given $Y_n(0) = i$ and $X(0) = x$;
 \[
p_i(t) = \E\left[\left.\mathbbm{1}_{Y_n(t)=i}\right\vert Y_n(0) = i, X(0) = x\right] \ \text{for} \ i \in \{S,I,V,R\}.
\]
The rates of change of these probabilities are as follows.
 \begin{subequations} \label{eq:prob_ode}
\begin{align}
\frac{\der p_S(t)}{\der t} & =   -N\kappa\alpha_S(t)\alpha_I(t) p_S(t) \E(X_I(t)) - \beta(t) p_S(t) \label{drift_pS}    \\
\frac{\der p_I(t)}{\der t} & =   N \kappa\alpha_S(t)\alpha_I(t)\E(X_I(t))  - \gamma(t) p_I(t)  \label{drift_pI}. \\
\frac{\der p_V(t)}{\der t} & =   \beta(t) p_S(t) + \gamma(t) p_I(t). 
\label{drift_pV}
\end{align}
\end{subequations}
 
Suppose $Y_n(0) = i$ and $X(0) = x$. Further, suppose the $n$th node employs an action function $\bar{u}(t) = (\bar{\alpha}(t), \bar{\beta}(t), \bar{\gamma}(t)), t \geq 0$ whereas all other nodes use $u(t) = (\alpha(t), \beta(t), \gamma(t)), t \geq 0$. Then the expected total cost of the tagged node is given by the following expression.
\begin{align*}
& \mathbb{E}\left[\right.\int_0^T \left(c_L(\bar{\alpha}_S(t)) \mathbbm{1}_{Y_n(t) = S} +  (c_L(\bar{\alpha}_I(t))+c_I(\bar{\gamma}(t))) \mathbbm{1}_{Y_n(t) = I}\right) \der t  \\
& \qquad  \qquad +  c_V\mathbbm{1}_{Y_n(T) = V}\ \vert \ Y_n(0) = i, X(0) = x\left.\right] \\
    &= \int_0^T \left(c_L(\bar{\alpha}_S(t))p_S(t)+  (c_L(\bar{\alpha}_I(t))+c_I(\bar{\gamma}(t)) p^N_I(t)\right) \der t+  c_V p_V(T).
\end{align*}

\subsection{The Case of Strategic Agents: Stochastic Game Model}
 \label{sec:stoch-game-prob}
 We now consider the individual nodes of the population to be strategic agents who wish to minimize their respective costs. These autonomous nodes also control their respective parameters giving rise to a stochastic game. We seek a symmetric Nash equilibrium, i.e., an equilibrium in which all the nodes employ the same policy. 
 
 We can formally describe a symmetric Nash equilibrium as follows. Suppose a tagged node uses a policy $\bar{\pi} = (\bar{\pi}_t, 0 \leq t \leq T)$ whereas all other nodes use a policy $\pi = (\pi_t, 0 \leq t \leq T)$. Then the occupancy measure of the other $N-1$ nodes is characterised by \eqref{eq:prob_ode}. 
 On the other hand, the tagged node's state evolves as \eqref{eq:prob_ode}
 with $u(t), t \in [0,T]$ replaced by $\bar{\pi}_t(X(t)), t \in [0,T]$, and its cost is 
 \begin{align*}
  \bar{J}_{\bar{\pi}}(i,x,\pi) & =  \int_0^T \left(c_L(\bar{\alpha}_S(t)) p_S(t)+ (c_L(\bar{\alpha}_I(t))+c_I(\bar{\gamma}(t))) p^N_I(t)\right) \der t \\ 
  & \quad + c_V p_S(T)
  \end{align*}
 where $(\bar{\alpha}(t),\bar{\beta}(t),\bar{\gamma}(t)) = \bar{\pi}_t(X(t))$.  A policy     
 $\bar{\pi}$ is called a symmetric NE if
\begin{equation} \label{eq:NE_strategy}
     \bar{\pi} \in \argmin_{\pi' \in  \Pi}\bar{J}_{\pi'}(i,x,\bar{\pi})
\end{equation}
 for all $x \in \Delta^N_4$. Nash equilibrium computation and implementation are marred with  computational complexity issues, much  as those faced by the computation of an optimal policy. 
The computational complexity problem  can be addressed by appealing to the mean field limit as described in Section \ref{sec3}. 

\section{Mean Field Modeling of Epidemic Spread} \label{sec3}
This section discusses mean field modeling of epidemics. The spread of epidemics can be modeled as the mean field limit of a sequence of dynamical processes. To present the key results in mean field analysis, for the sake of convenience, we move away from the SIVR model presented in the previous section to a more general $K$-compartmental model. We present the standard solution to the mean field control problem. We then consider the strategic case and present the mean field game model and a solution for the same.

We start our discussion by considering  a sequence of systems with increasing population sizes. Under certain regularity conditions to be described below, the corresponding sequences of processes $X^N(t), t \geq 0$ and $p^N(t), t \geq 0$ converge to deterministic functions referred to as mean field limits. Below we discuss this convergence in the context of a more general dynamical system; spread of epidemics in a population will be a special case.     

\subsection{Mean Field Limit in a General Dynamical System}
Let us consider a population of $N$ nodes, where each node can be in one of the $K$ states, $\{1,\ldots,K\} =: \mathcal{S}$. Let $Y^N(t) = (Y^N_1(t),\ldots,Y^N_N(t))$  denote the states of the nodes at time $t$ as in Section \ref{sec2.2}. 

\subsubsection{Evolution of Population in the Mean Field Case}
Let $\pi^N = (\pi^N_{i,t}, i \in \mathcal{S}, t \in [0,T])$ where $\pi^N_{i,t}: \Delta^N_K \to U_i$ for all $t \in [0,T]$ and $U_i$ are compact subsets of  $\mathbb{R}_+^{L_i}$, be the policy used by each of the nodes.
Further, with the occupancy measure $X^N(t) \in \Delta^N_K, t \geq 0$ as defined in Section \ref{sec2.2},  the expected conditional drift rates are, for $j = 1,\ldots,K$, 
\[
\frac{\der \mathbb{E}[X^N_j(t) |X^N(t)=x]}{\der t} = \sum_{i=1, i\neq j}^K x_i Q_{ij}(x_j,\pi^N_{i,t}(x),\pi^N_{j,t}(x)) \ 
\]
 Here, for any $i, j~(j \neq i)$, given $X^N(t)=x$, $Q_{ij}(x_j,\pi^N_{i,t}(x),\pi^N_{j,t}(x))$ is the state transition rate of a node from state $i$ to state $j$, and 
 \[Q_{ii}(x,\pi^N_{i,t}(x),j=1,\dots,K) \coloneqq - \sum_{j \neq i}Q_{ij}(x_j,\pi^N_{i,t}(x),\pi^N_{j,t}(x)).
 \]
 We can also write a general form of average cost of the population, $J^N_{\pi^N}(x)$, as follows.
\begin{align*}
 J^N_{\pi^N}(x)  = \E\left[\right.\int_0^T X^N(t)^T g(X^N(t),\pi^N_{i,t}(X^N(t)),i \in \mathcal{S}) \der t  \\
  +  X^N(T)^T h(X^N(T)) \ \vert \ X^N(0) = x\left.\right]
\end{align*}
where $g(x,\pi^N_{i,t}(x),i \in \mathcal{S}) \coloneqq (g_i(x,\pi^N_{i,t}(x)),i=1,\dots,K)$ are the cost rates and $h(x) \coloneqq (h_1(x), \cdots,h_K(x))$ are the terminal costs.
The optimal control problem and its solution are as in \eqref{eq:value_pol_opt}.
 
\subsubsection{Evolution of an Individual Node in the Mean Field Case} 
Consider a tagged node, say the $n$th node, which, at time $t=0$ is in state $i \in \mathcal{S}$ and the initial population state, $X^N(0) = x$. Further, suppose that this node employs a policy $\bar{\pi}^N = (\bar{\pi}^N_{i,t}, i \in \mathcal{S}, t \in [0, T])$ whereas all other nodes use $\bar{\pi}^N = (\bar{\pi}^N_{i,t}, i\in \mathcal{S}, t \in [0, T])$. Let $p^N(t) \coloneqq (p^N_1(t), \cdots,p^N_K(t))$ denote the conditional probabilities of the $n$th node being in states $1,\cdots,K$, respectively, at time $t$.
The rate of change of these probabilities can be expressed as follows.
\begin{align*} 
\frac{\der p^N_j(t)}{\der t} = \sum_{i=1,i \neq j}^K p^N_i(t) \E[Q_{ij}(X^N_j(t),\bar{\pi}^N_{i,t}(X^N(t)),\bar{\pi}^N_{j,t}(X^N(t)))] \\ \text{ for } j = 1,\cdots,K.  
\end{align*}
Further, the $n$th node's cost, $\bar{J}_{\bar{\pi}^N}(i,x,\pi^N)$, can be expressed as follows.
\begin{align*}
\bar{J}^N_{\bar{\pi}^N}(i,x,\pi^N) = \int_0^T p^N(t)^T g(X^N(t),\pi^N_{j,t}(X^N(t)),j \in \mathcal{S}) \der t \\
+ p^N(T)^T h(X^N(t)).
\end{align*}
Finally, the Nash equilibrium can be characterised as in \eqref{eq:NE_strategy}. 

\paragraph{Example~(Epidemics):} 
In the special case of evolution of epidemics as in Section \ref{sec2}, $K = 4, L_1 = L_2 = 3, L_3 = L_4 = 0$ and for $u_1=(\alpha_S,\beta)$ and $u_2=(\alpha_I,\gamma)$, 
\[
Q(x,u_1,u_2) = \begin{bmatrix}
-N\kappa \alpha_S \alpha_I x_I - \beta & N\kappa\alpha_S \alpha_I x_I & \beta & 0\\
0 & - \gamma&  0 & \gamma \\
0 & 0 & 0 & 0\\
0 & 0 & 0 & 0.
\end{bmatrix}.
\]
Moreover, 
\begin{align*}
g(x,u_1,u_2) & = [c_L(\alpha_S) \ \ c_L(\alpha_I)+c_I(\gamma) \ \ 0 \ \ 0]^T    \\
h(x) &= [0 \ \ 0 \ \ 0 \ \ c_V]^T.
\end{align*}

\subsection{Mean Field Control}
Darling~\cite{darling2002fluid} analyse the convergence of CTMCs to solutions of certain ODEs. Gast et al. \cite{GAST12} study convergence of controlled DTMCs to continuous times controlled deterministic dynamical systems. Following these works, we make the following hypotheses on the dynamics and the costs. 

\subsubsection{Mean Field Convergence}
\label{sec:MFconvergence}

\paragraph{Assumptions on dynamics}
\begin{enumerate}
    \item {\it Initial Conditions:} The initial occupancy measure $x^N(0)$ converges to $x(0) \in \Delta_K$ in probability, i.e.,
    $\lim_{N\to\infty}P(\Vert X^N(0)-x(0) \Vert > \epsilon) = 0$ for all $\epsilon > 0$.
    \item {\it Transition rates:} For all $i,j~(j \neq i)$, the state transition rates $Q_{ij}(x_j,u_i,u_j)$ are  $O(1)$. This implies that the expected number of transitions per unit time is $O(N)$.
    \item {\it Drift rates:} The drifts are $O(N)$. Moreover, there exist bounded transition rate matrices $Q(\cdot,u) \in \mathbb{R}^{K\times K}$ $ \forall u \in U \coloneqq \bigotimes_{i=1}^K U_i$, such that $x^T Q(x,u)$ converges to $x^T Q(x,u)$ uniformly in $(x,u)$.  
    \item {\it Lipschitz continuity:}  $x^T Q(x,u)$  is Lipschitz continuous in $(x,u)$.
\end{enumerate}

\paragraph{Assumptions on the costs}
\begin{enumerate}
\item {\it Cost rates:}
The cost rates $g_i(\cdot,u_i)$ and the terminal costs $h_i(\cdot)$ are bounded.
\item {\it Lipschitz continuity:}
The cost rates $g_i(\cdot,u)$ are Lipschitz continuous for all $u_i \in U_i$, and the terminal costs $h_i(\cdot)$ are also  Lipschitz continuous.
\end{enumerate}

Observe that the scaling conditions are satisfied in the case of epidemics if we choose $\kappa^N = \kappa/N, \alpha_S^N = \alpha_S,$$ \alpha_I^N = \alpha_I, \beta^N = \beta$ and $\gamma^N = \gamma$. We further assume that $c_L(\cdot)$ and $c_I(\cdot)$ are bounded functions. It can be easily checked that the above hypotheses are satisfied under these assumptions. 

In order to state the mean field convergence results, we introduce the following optimal control problem. Consider a continuous time dynamical system
\begin{equation} \label{eq:state_dyn}
\frac{\der x(t)}{\der t}  = Q^T(x(t),u(t))x(t) 
\end{equation}
with state $x(t), t \geq 0$, initial condition $x(0)$, and action function $u(t) = (u_i(t),i=1,\cdots,K) , t \geq 0$. Let $J_u: \Delta_K \to \mathbb{R}_+$  be the associated cost function defined as follows.   
\begin{equation*}
J_u(x) = \int_0^T x(t)^T g(x(t),u(t))\der t +  x(T)^T h(x(T))
\end{equation*}
where $x(t), t \geq 0$ is a solution to~\eqref{eq:state_dyn} given $x(0) = x$. 
We seek to find an action function $u^\ast(t), 0 \leq t \leq T$ which together with its corresponding state trajectory $x^\ast(t), 0 \leq t \leq T$ attains the minimum cost
\[
J(x) \coloneqq \min_{u} J_u(x).
\]
We then have the following convergence results. 
\begin{enumerate}
\item {\it Optimal trajectory:} The CTMC $X^N(t), 0 \leq t \leq T$ converges to the mean field limit $x(t), 0 \leq t \leq T$. More precisely, for all $\epsilon >0$, 
\[
\lim_{N \to \infty}\mathbb{P}\left[\sup_{0 \leq t \leq T} \Vert X^N(t)- x(t)\Vert > \epsilon \right] = 0
\]
\item {\it Optimal cost:} The optimal cost for the population of size $N$, $J^N(X^N(0))$, converges to the optimal cost of the mean field control problem, $J(x(0))$, in probability.
\item {\it Optimal policy:} An optimal action function for
the mean field limit, say $u^\ast$,  is asymptotically optimal for the system with finitely many nodes. More precisely, $J^N_{\pi^\ast}(X^N(0)) -  J(x(0))$ converges to $0$ in probability
where $\pi^\ast_{i,t} \equiv u^\ast_i(t)$ for all $i$. 
\end{enumerate}

\subsubsection{Solution to the Mean Field Control Problem}
\label{sec:MFC_soln}
With a slight abuse of notation, we let $J_t(x)$ denote the optimal cost over $[t,T]$ given $x(t) = x$. We can write
\begin{align*}
J_t(x) &= \min_{u \in U}\left\{x^Tg(x,u)\der t+J_{t+\der t}(x(t+\der t))\right\} \\ 
&= \min_{u \in U}\left\{x^T g(x,u)\der t + J_{t+\der t}(x) + x^T Q(x,u) \nabla_x J_t(x) \der t\right\}
\end{align*} 
On rearranging the terms, we obtain
\[
-\nabla_t J_t(x) = \min_{u \in U} \left\{x^T g(x,u) + x^T Q(x,u) \nabla_x J_t(x)\right\}.
\]
The above equation, referred to as the Hamilton-Jacobi-Bellman~(HJB) equation, yields both the optimal action function $u^\ast: [0,T] \to U$ and the optimal cost $J(x) \equiv J_0(x)$ of the mean field control problem.

\paragraph{The Minimum Principle \cite{bertsekas2012dynamic} :}
Let us introduce adjoint~(also called {\it costate}) processes $\lambda(t) \in \mathbb{R}^K, t \in [0,T]$ and the Hamiltonian function $H: \Delta^N_K \times U \times \mathbb{R}^K \to \mathbb{R}$ given by
\[
H(x,u,\lambda) =  x^T g(x,u) + x^T Q(x,u)\lambda.
\]
Let  $u^\ast(t), t \in [0,T]$ be an optimal action function and $x^\ast(t), t \in [0,T]$ be the corresponding state trajectory, i.e., $x^\ast(t), t \in [0,T]$ is the solution to~\eqref{eq:state_dyn} with $u(t) = u^\ast(t), t \in [0,T]$ and initial condition $x^\ast(0) = x$.  The HJB equation can be compactly written in terms of the Hamiltonian function as
\[
-\nabla_t J_t(x^\ast(t)) = H(x^\ast(t),u^\ast(t),\nabla_x J_t(x^\ast(t))).
\]
The following result, referred to as {\em the Minimum Principle\/}, provides a necessary condition for optimality of $u^\ast(t), t \in [0,T]$.
Let $\lambda(t), t \in [0,T]$ be the solution to equation
\[
\frac{\der \lambda(t)} {\der t} = -\nabla_x H(x^\ast(t),u^\ast(t),\lambda(t)),   
\]
referred to as the adjoint equation, with the boundary condition: 
\[
\lambda(T) = \nabla (x^\ast(T)^T h(x^\ast(T))).  
\]
Then, for all $t \in [0,T]$,
\[
u^\ast(t) = \argmin_{u \in U} H(x^\ast(t),u,\lambda(t)).   
\]
Furthermore, there is a constant $\theta$ such that
\[
H(x^\ast(t),u^\ast(t),\lambda(t)) = \theta, \ \text{for all} \ t \in [0,T].   
\]
The minimum principle can be used as the basis of a numerical solution. In the so called {\it two-point boundary problem method}, we use the necessary condition
\[
u^\ast(t) = \argmin_{u \in U} H(x^\ast(t),u,\lambda(t)).   
\]
to express $u^\ast(t)$ in terms of $x^\ast(t)$ and $\lambda(t)$. We then substitute the result into the system and the adjoint equations, to obtain a set of $2K$ first order differential equations in $x^\ast(t)$ and $\lambda(t)$. These equations can be solved using the split boundary conditions: 
\[x^\ast(0) = x \ \text{and} \ \lambda(T) = \nabla (x^\ast(T)^T h(x^\ast(T))).\]

\subsubsection{Solution to the Finite Population Control Problem}
We thus obtain the following procedure for solving the optimal control problem for a finite population.
\begin{itemize}
    \item From the original system with $N$ nodes, write the mean field limit. In particular, set $x(0)= X^N(0)$ and obtain $Q(\cdot,u)$ via appropriate scaling of parameters.   
    \item Obtain the optimal control for the limiting problem via solving the HJB equation or via some other method, e.g., using the minimum principle \cite{bertsekas2012dynamic}.
    \item Use this control in the finite population problem. From the above discussion, this is asymptotically optimal. 
\end{itemize}

\subsection{Mean Field Game Model of Epidemic Spread}
\label{sec:mfg}
Let us focus on a tagged node, say the $n$th node. Suppose it uses an action function $\bar{u}: [0,T] \to U$ whereas all other nodes use an action function $u: [0,T] \to U$. Following the discussion in Section \ref{sec:MFconvergence}, the occupancy measures of these nodes, $X^N(t), t \in [0,T]$ asymptotically approach the mean field limit $x(t), t \in [0, T]$. Hence, in the limiting system, the probabilities of the tagged node being in various states, $p(t), t \in [0, T]$, given its initial state, say $i$, evolve as follows. 
\begin{equation} \label{eq:indiv_dyn}
\frac{\der p}{\der t}  = Q^T(x(t),\bar{u}(t)) p(t) 
\end{equation}
with $p(0) = \delta_i$. Let $\bar{J}_{\bar{u}}(i,x,u)$  be the associated cost defined as follows. 
\begin{equation*}
\bar{J}_{\bar{u}}(i,x,u) = \int_0^T p(t)^T g(x(t),u(t))\der t +  p(T)^T h(x(T))
\end{equation*}
where $x(t), 0 \leq t \leq T$ and $p(t), 0 \leq t \leq T$ are solutions to \eqref{eq:state_dyn} and \eqref{eq:indiv_dyn},  respectively, given $x(0) = x$ and $p(0) = \delta_{i}$. We seek to find a action function $\bar{u}: [0,T] \to U$, which, together with the corresponding trajectory $\bar{p}(t),  t \in [0, T]$, 
minimizes the tagged node's cost when other nodes are also using the same action function $\bar{u}:[0, T] \to U$, i.e.,
\[
\bar{u} \in \argmin_{u'} \bar{J}_{u'}(i,x,\bar{u}).
\]
We expect the equilibrium trajectory, policy, and cost for the stochastic game for a finite size population to be asymptotically close to those for the mean field game. This is a conjecture and we do not yet have a  proof of convergence.

\subsubsection{Solution to the Mean Field Game}
\label{sec:mean_field_solution}
With a slight abuse of notation we let $\bar{J}_t(i,x,u)$ denote the optimal cost of the $n$th node over $[t,T]$ given $Y_n(t) = i, X(0) = x$. We can write
\begin{align*}
\bar{J}_t(i,x,u) = \min_{u'_i \in U_i}&\mathbb{E}\left[\right.g_i(x(t),u'_i)\der t+\bar{J}_{t+\der t}(Y_n(t+\der t), x,u) \\
& \left. \ | \ Y_n(t) = i, x(0) = x\right] \\ 
=\min_{u'_i \in U_i}&\left\{g_i(x(t),u'_i)\der t + \sum_{j \neq i} Q_{ij}(x_j(t),u'_i,u_j(t))\der t \bar{J}_{t+\der t}(j,x,u)\right. \\
&\left.+ \left(1 - \sum_{j \neq i} Q_{ij}(x_j(t),u'_i,u_j(t))\der t\right) \bar{J}_{t+\der t}(i,x,u)\right\}
\end{align*} 
On rearranging the terms we obtain
\begin{align} 
-\nabla_t \bar{J}_t(i,x,u) &= \min_{u'_i \in U_i} \left\{\right.g_i(x(t),u'_i) \nonumber \\ 
& + \sum_{j \neq i} Q_{ij}(x_j(t),u'_i,u_j(t))(\bar{J}_t(j,x,u)-\bar{J}_t(i,x,u))\left.\right\} \nonumber \\
&= \min_{u'_i \in U} \left\{g_i(x(t),u'_i) + \sum_{j} Q_{ij}(x_j(t),u'_i,u_j(t))\bar{J}_t(j,x,u)\right\}.\label{eq:HJB}
\end{align}
This is the HJB equation for the mean field game. A control $\bar{u}$ is a Nash equilibrium of the game provided 
\[
\bar{u}_i(t) \in \argmin_{u'_i \in U_i} \left\{g_i(x(t),u'_i) + \sum_j Q_{ij}(x_j(t),u'_i,\bar{u}_j(t)))\bar{J}_t(j,x,\bar{u})\right\}
\]
for all $t \in [0,T]$. So, Nash equilibria are characterized by the system of Kolmogorov and HJB equations, \eqref{eq:state_dyn} and \eqref{eq:HJB}, respectively, together with the boundary conditions:
\begin{align*}
x(0) &= x\\ 
\text{and } \bar{J}_T(i,x,\bar{u}) &= h_i(x(T)).
\end{align*}
Thus we get a initial-terminal value problem~(ITVP) whose fixed points yield the solutions to the mean field game.
Under regularity assumptions for the cost functions $g_i(\cdot,\cdot)$ and the transition rates $Q(\cdot,\cdot)$, it can be shown that a unique minimizer exists in \eqref{eq:HJB}, which is the Nash equilibrium of the game \cite[Section 7.2.2]{CARMONA18}. 
Following is a fixed point iteration to obtain a Nash equilibrium given $x(0)$.
\begin{itemize}
    \item Initialize with an action function $u: [0, T] \to U$.
    \item Solve Kolmogorov equations \eqref{eq:state_dyn} to obtain trajectory $x(t), 0 \leq t \leq T$ corresponding to $u$.
    \item For each $j \in \{1,\cdots,K\}$, solve HJB equations \eqref{eq:HJB} to obtain the best responses $
    u'_i: [0,T] \to U_i$ to  $u$ for all $i$. 
    \item If $u'(t) = u(t)$ for all $t \in [0,T]$, set $\bar{u} = u'$ and exit. Otherwise, set $u = u'$, and continue with the next iteration~(go to Step~2). 
\end{itemize}

Let $f_x$ denote the mapping from $u$ to $\bar{u}$. When $f_x$ is a \emph{contractive} mapping on a complete metric space, Picard-Banach fixed point theorem states that there is a unique fixed point and that the convergence is geometric \cite{BREMAUD21}.

\textbf{The Master Equation:} \cite[Section 6.5]{CARMONA18}
Let us introduce the adjoint processes $\bar{\lambda}(t) \in \mathbb{R}^K, t \in [0,T]$ and the Hamiltonian functions $H_i: \Delta^N_K \times U \times \mathbb{R}^K \to \mathbb{R}$ given by
\[
H_i(x,u,\lambda) =  g_i(x,u_i) + \sum_{j}Q_{ij}(x_j,u_i,u_j) \lambda_j.
\]
Let  $\bar{u}_i(t), i \in \mathcal{S},t \in [0,T]$ be the best response action functions of the tagged player. The HJB equations can be compactly written in terms of the Hamiltonian functions as
\begin{equation*}
-\nabla_t \bar{J}_t(i,x,u) = 
\min_{u'_i \in U_i} H_i(x(t),(u'_i,u_{-i}(t)),\bar{J}_t(\cdot,x,u)) \ \forall i \in \mathcal{S}, 
\end{equation*}
and
\begin{equation*}
\bar{u}_i(t) \in \argmin_{u'_i \in U_i} H_i(x(t),(u'_i,\bar{u}_{-i}(t)),\bar{J}_t(\cdot,x,\bar{u})) \ \forall i \in \mathcal{S}, 
\end{equation*}
where $\bar{J}_t(\cdot,x,u) \in \mathbb{R}^K$ has $j$th component $\bar{J}_t(j,x,u)$ for all $j \in \mathcal{S}$. Finally, let ${\cal J}_t, t \in [0,T]$ be real valued functions, ${\cal J}_t: [K] \times \mathbb{R}^K \to \mathbb{R}$, and let us also introduce the following equation known as the master equation.
\begin{align}
\nabla_t{\cal J}_t(i,x) + H_i(x,(\nu_i(x,u_{-i}(t),{\cal J}_t(\cdot,x)),u_{-i}(t)),{\cal J}_t(\cdot,x))  \nonumber \\
+ \sum_j \left(\sum_l 
x_l Q_{lj}(x_j,\nu_i(x,u_{-i}(t),{\cal J}_t(\cdot,x)),u_j(t))\right)
\nabla_{x_j}{\cal J}_t(i,x)) = 0 
\end{align}
where
\[
\nu_i(x,u_{-i},\lambda) \in \argmin_{u'_i \in U_i} H_i(x,(u'_i,u_{-i}),\lambda)
\]
where
\[
q_j(x,u,z) = \sum_l x_l Q_{lj}(x_j,u_l(z),u_j).  
\]
Let ${\cal J}_t, t \in [0,T]$ be a solution to the master equation with terminal condition ${\cal J}_T(i,x) = h_i(x)$ for all $i \in \mathcal{S}$, and $\bar{u}(t), t \in [0,t]$ be the associated optimal action function. Let $\bar{x}(t), t \in [0,T]$ be a solution to 
\begin{equation}
\frac{\der \bar{x}_i(t)}{\der t} =  q_t(i,\bar{x}(t),\bar{u}(t),{\cal J}_t(\cdot,\bar{x}(t)))  \label{eqn:Kolmogorov2}
\end{equation}
with initial condition $\bar{x}(0)$. 
Then, setting $\bar{J}_t(i,x,u)  = {\cal J}_t(i,\bar{x}(t))) $,
\begin{align*}
&\nabla_t\bar{J}_t(i,x,u) + H_i(x(t),\bar{u}(t),\bar{J}_t(\cdot,x(t),u(t))) \\
& = \nabla_t{\cal J}_t(i,x) + \sum_j q_t(j,x,\bar{u}(t),{\cal J}_t(\cdot,x)) \nabla_{x_j}{\cal J}_t(\cdot,x)) \\
& \quad + H_i(x,\bar{u}(t),{\cal J}_t(\cdot,x)) \\
&= 0. 
\end{align*}
Clearly, $\bar{J}_t(i,x,u)$ solves the HJB equation and is value function in the optimization problem~\eqref{eq:HJB}. Also, \eqref{eqn:Kolmogorov2} can be identified with the Kolmogorov equation~\eqref{eq:state_dyn}. Consequently, $\bar{x}(t)$ is the equilibrium state trajectory. It is therefore seen that the master equation encapsulates both the Kolmogorov and the HJB equations in a single equation. 


\subsubsection{Solution to the Finite Population Stochastic Game}
Finally, we can adopt a procedure similar to that in Section \ref{sec2.2}, to obtain a solution to the stochastic game for a population of size $N$. 

\section{An Illustrative Example}
We will now study how to formulate a mean field control as well as a mean field game problem in a compartmental epidemic model.
For ease of  our presentation, we  simplify the SIVR model introduced in Section \ref{sec2} to an SIR model without vaccination.
We assume that the transmission rate of the disease, $\kappa$, is influenced by two factors: the disease characteristics and the baseline contact factor within the population. 
The recovery rate $\gamma$ is taken to be a constant.
Individuals can adjust  their level of social interaction by choosing their \emph{contact factor}. This is reflected by choosing a value of $\alpha(t) \coloneqq (\alpha_S(t), \alpha_I(t)) \in [\alpha_{\min}, 1] \times [\alpha_{\min}, 1] \eqqcolon A$. $\alpha_{min}$ represents the minimal transmission rate that an individual who adopts maximum protection effort will encounter.

\subsection{Illustrative Example: Mean Field Control}
From a societal viewpoint, we assume that the regulating authority chooses a policy $\pi =(\pi_t, t \in [0,T])$ for the whole population. This can be thought of as a combination of non-pharmaceutical interventions like lockdowns, mandatory usage of masks, proactive  testing and quarantine, etc.
We look at a finite horizon setup to identify the social optimum strategy that minimizes the total social cost.

Once the social planner commits to a policy $\pi(t)$, analogous to equation \eqref{eq:state_dyn}, the forward evolution of the state with the initial conditions $x(0)=(x_S(0),x_I(0),x_R(0))$
is given by:
\begin{subequations} \label{SIR_contact}
\begin{align}
\dot x_S(t) & =  -\kappa\alpha_I(t)\alpha_S(t) x_S(t) x_I(t)  \\
\dot x_I(t) & =  \kappa\alpha_I(t)\alpha_S(t) x_S(t) x_I(t)-\gamma x_I(t) \\
\dot x_R(t) & =   \gamma x_I(t)
\end{align}
\end{subequations}
We assume that individuals in states $i \in \{S,I\}$ choosing a contact factor $\alpha_i(t)$ incurs a running cost $c_L(\alpha_i(t)) = c_Q (1-\alpha_i(t))$. In addition to this lockdown cost, infected individuals pay a cost $c_I$ per unit time.

Starting from $x(0)=(x_S(0),x_I(0),x_R(0))$, policy $\pi$ induces a social cost which can be expressed as: 
\begin{eqnarray*}\label{cost_opt}
J_{\pi}(x)=\int_0^{T} \left( c_Q (1-\alpha_S(t)) x_S(t) + (c_I+c_Q(1-\alpha_I(t))) x_I(t) \right)\der t
\end{eqnarray*}
We can define the Hamiltonian corresponding to the optimal control problem as, 
\begin{align*}
H(x,\pi,\lambda) &= c_Q (1-\alpha_S(t)) x_S(t) + (c_I+c_Q (1-\alpha_I(t))) x_I(t) \\ & \ - \lambda_2(t) \gamma x_I(t) 
  + \kappa\alpha_I(t)\alpha_S(t) x_S(t) x_I(t) (\lambda_2(t) -\lambda_1(t)) 
\end{align*}
Defining $\lambda(t)=(\lambda_1(t),\lambda_2(t),\lambda_3(t))$ as the co-state variable vector, with $\lambda_3(t):=0 \quad \forall t \in [0,T]$,

\begin{align}\label{costate_ode}
\dot \lambda_1  & = -\dfrac{\partial H}{\partial x_S}  =   - c_Q (1-\alpha_S(t))+ \kappa \alpha_I(t) \alpha_S(t) x_I(t) (\lambda_1(t) -\lambda_2(t))  \nonumber \\
\dot \lambda_2  &=  -\dfrac{\partial H}{\partial x_I} =  -c_I - c_Q (1-\alpha_I(t))+ \lambda_2(t) \gamma \nonumber \\
& \qquad \qquad + \kappa \alpha_I(t) \alpha_S(t) x_S(t) (\lambda_1(t) -\lambda_2(t)) 
\end{align}
The final constraints for the co-state variables are: $\lambda_1(T)=\lambda_2(T)=0$.

From our discussion in section \ref{sec:MFC_soln}, functions $\lambda_1, \lambda_2$ exist that satisfy equation \eqref{costate_ode}, and the optimal control $\pi^{*}$ is the minimization of Hamiltonian, assuming that the co-state variables are set according to optimum control. 

\subsubsection{Numerical Computation of Optimal Control}
Note that in the previous section, we had to deal with two sets of ODEs. We have the forward evolution ODEs of the compartmental model \eqref{SIR_contact} along with their initial conditions. In addition to which, we had the co-state ODE \eqref{costate_ode} with terminal conditions. There are several classical procedures available to compute the optimal control. 

\begin{figure}[h!]
        \centering
    \includegraphics[width=0.5\textwidth]{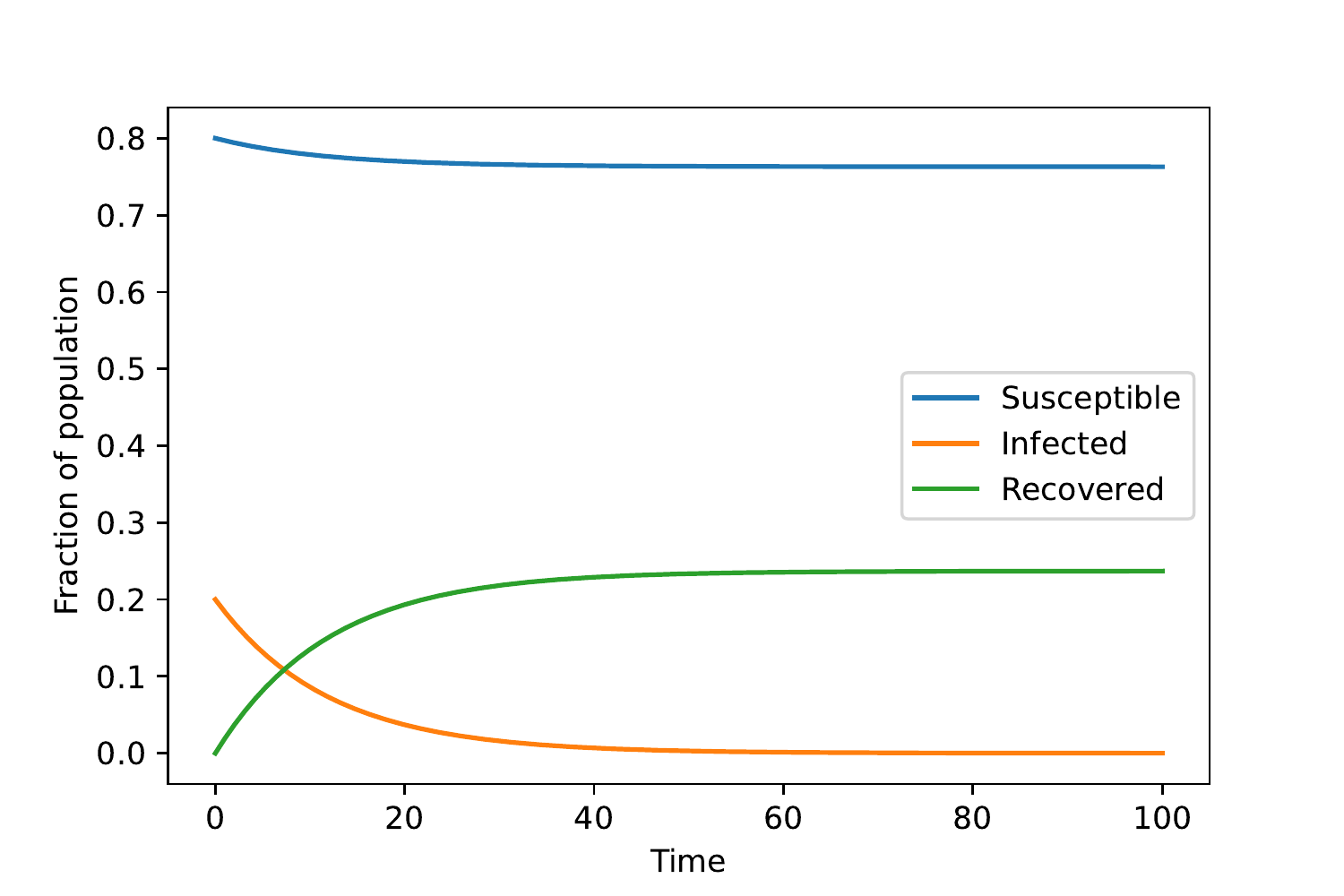}
    \caption{Time evolution of the state $(x_S,x_I,x_R)$ under socially optimal control policy.}
    \label{fig:ode_evolution}
\end{figure}

\begin{figure}[h!]
    \centering
    \includegraphics[width=0.5\textwidth]{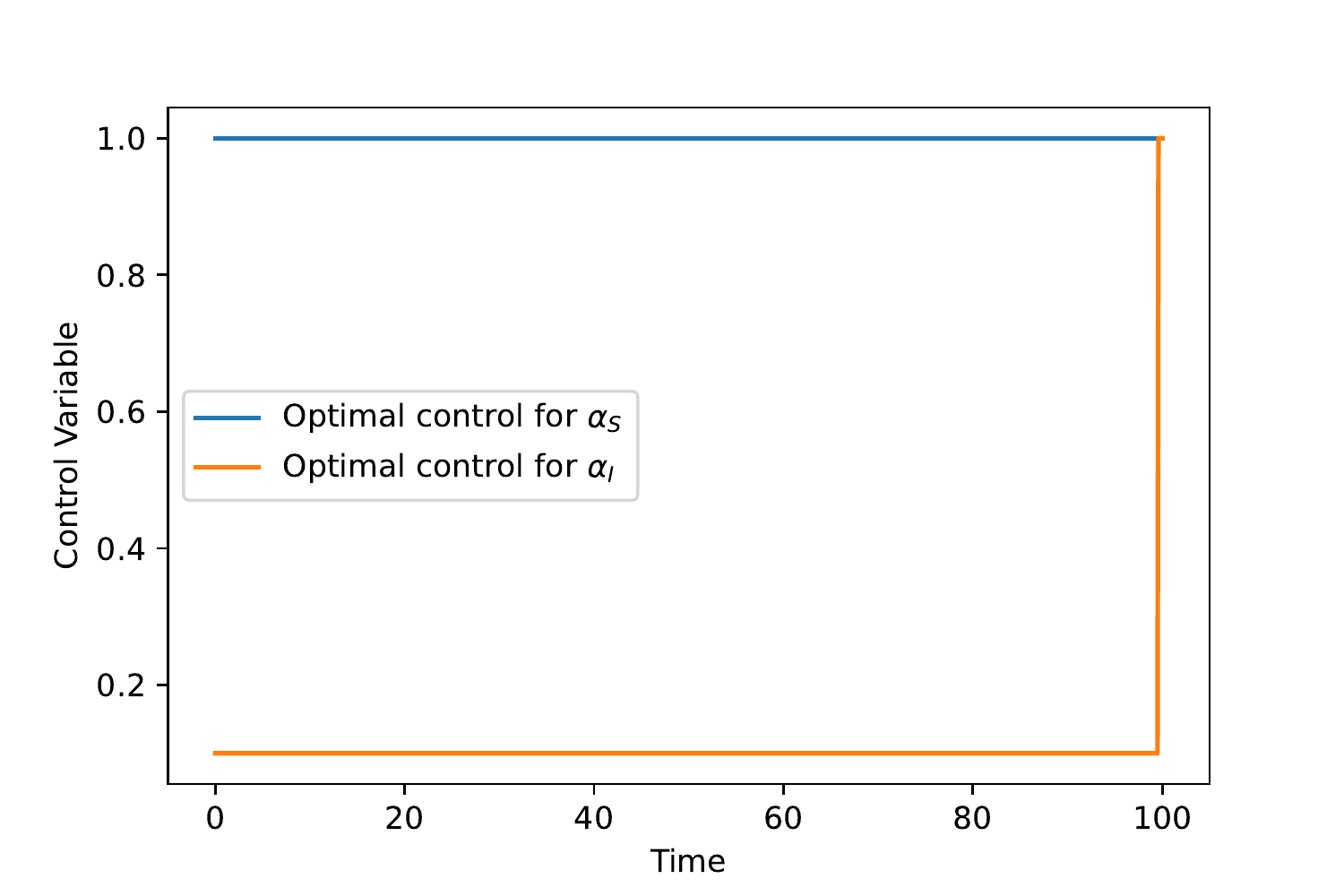}
    \caption{Socially optimal control policy.}
    \label{fig:opt_Law}
\end{figure}

We use the forward-backward sweep method in \cite[Chapter 4]{lenhart2007optimal}. 
Figure \ref{fig:ode_evolution} shows time evolution of the epidemic in the population. Figure \ref{fig:opt_Law} shows the socially optimal control for the set of parameters in table \ref{tab:parameters}.

\begin{table}[h!]
    \centering
    \begin{tabular}{|c|c|c|c|c|c|c|c|c|c|}
        \hline
        $(S_0,I_0,R_0)$ & $\gamma$ & $\kappa$  & $T$ & $c_I$ & $c_Q$ & $\alpha_{min}$  \\
        \hline
        $(0.8,0.2,0)$&  $0.1$ & $0.2$ & $100$ & $100$ & $10$  &  $0.1$ \\
        \hline
    \end{tabular}
    \caption{Set of parameters for the numerical experiments}
    \label{tab:parameters}
\end{table}

\subsection{Illustrative Example: Mean Field Game Model}
From the point of view of a strategic  individual node, computing the best response to a population strategy can be seen as a mean field game. 
Suppose a tagged node uses a policy $\bar{u}=(\bar{\alpha}(t))_{t \in [0,T]}$ whereas all other nodes use a policy $u$. The cost of an individual node is
 \begin{align*}
  \bar{J}_{\bar{u}}(i,x,u) & =  \int_0^T   \{c_Q (1-\bar{\alpha}_S(t)) p_S(t) \\
  & \qquad + (c_I+c_Q(1-\bar{\alpha}_I(t))) p_I(t)\}  \der t.
 \end{align*}
We seek to find a policy $\bar{u}: [0,T] \to U$ which, together with the corresponding trajectory $\bar{p}(t),  t \in [0, T]$ 
minimizes the tagged node's cost when other nodes are also using the same policy $\bar{u}$ \eqref{eq:NE_strategy}.

Writing the corresponding Bellman equations (from section \ref{sec:mfg}), with $J_i(t)$ denoting the cost to go at time $t$ from state $i \in \{S,I\}$ we have:
\begin{subequations} \label{eq:bellman_example}
\begin{align}
    -\dot J_S(t) & =  \min_{\alpha_S(t)} c_Q (1-\alpha_S(t)) + \kappa \alpha_S(t) \alpha_I(t) p_I(t) (J_I(t)-J_S(t)) \\
    -\dot J_I(t) & =  \min_{\alpha_I(t)} c_I + c_Q (1-\alpha_I(t)) - \gamma J_I(t) 
\end{align}
\end{subequations}
along with the terminal conditions $J_S(T)=J_I(T)=0$. The cost-to-go from recovered state $J_R(t)$ is identically set to zero. From \eqref{eq:bellman_example}, we can immediately see that for an infected individual, $\alpha_I(t)=1$ is the best response action. Defining a switching function $\phi(t) := \kappa p_I(t) (J_I(t)-J_S(t)) - c_Q$, we can find the individual best response strategy for a susceptible individual as follows:
for any $t \in [0,T] $, 
\begin{eqnarray}
\alpha_S^{*}(t)=
\begin{cases}
\alpha_{min} & \text{ if } \phi(t) \geq 0 \\
1 & \text{ if } \phi(t) < 0 
\end{cases}
\nonumber
\end{eqnarray}

\subsubsection{Numerical Computation of Mean Field Equilibrium}

\begin{figure}[h!]
    \centering
    \includegraphics[width=0.5\textwidth]{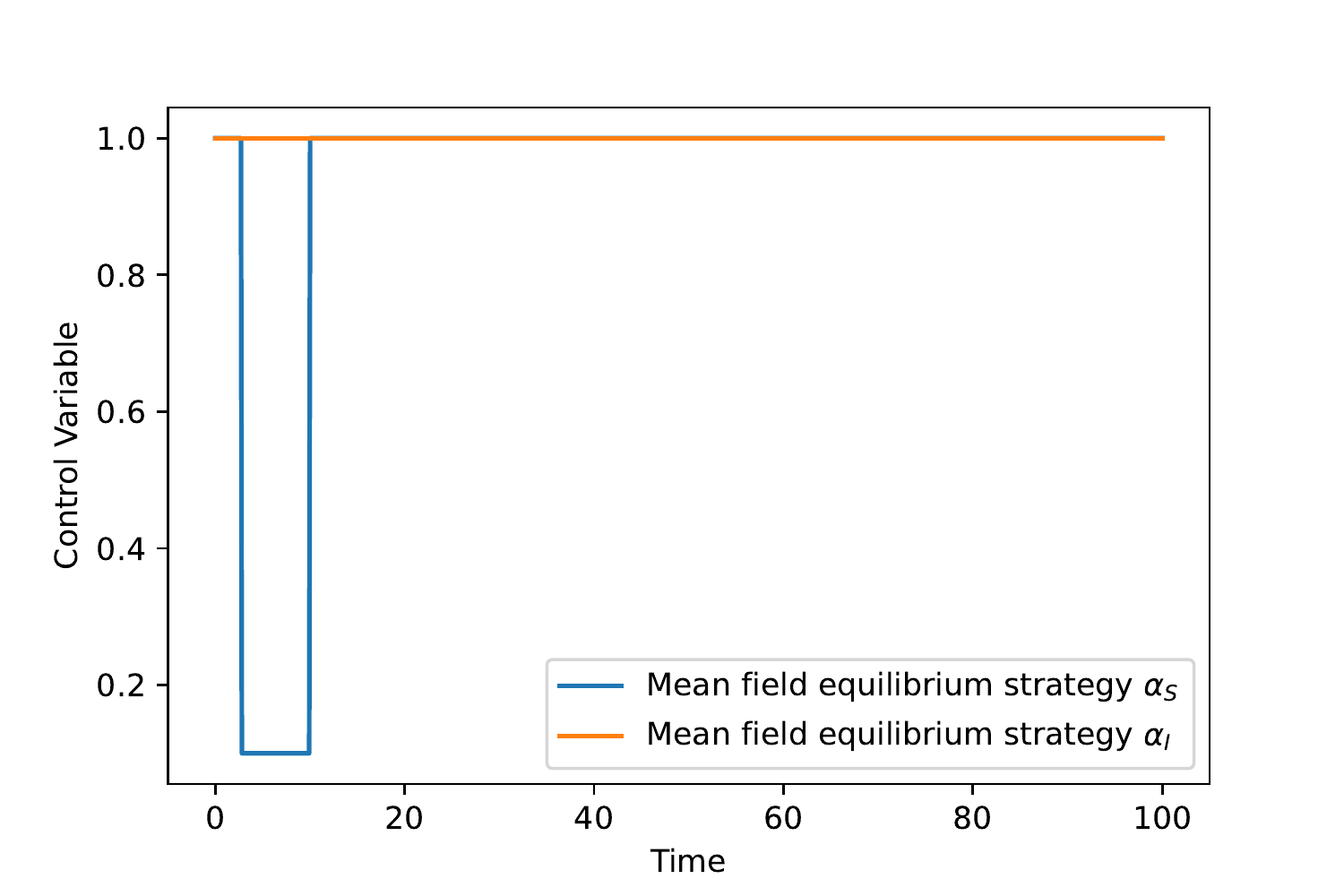}
    \caption{Mean field equilibrium control policy.}
    \label{fig:mfg_Law}
\end{figure}

Nash equilibria correspond to a fixed point of the best response function. The idea is to use an inductive sequence $u_{n+1} = f_{x_0}(u_n)$ to reach the fixed point. The Kolmogorov equations \eqref{eq:state_dyn} are solved for a given control $u_n$ and the best response is computed from solving the Bellman ODE \eqref{eq:bellman_example} backward in time, giving $u_{n+1}$. We use the iterative scheme detailed in section \ref{sec:mean_field_solution} to compute the mean field equilibrium.

Figure \ref{fig:mfg_Law} shows the individually optimal control for the set of parameters in Table \ref{tab:parameters}. We observe that the socially optimal control policy requires that the population should reduce their social interaction for a longer period compared to the mean field equilibrium policy.

\subsection{Comparison of Mean Field Control and Mean Field Equilibrium strategies }

We can see from figure \ref{fig:opt_Law} that in an idealized setting where the regulating authority has full control on contact between individuals, only the infected needs to be isolated and this is reflected in the control actions $\alpha_S$ and $\alpha_I$ chosen for individuals in Susceptible and Infected states respectively.
However, when we look at a mean field equilibrium strategy (figure \ref{fig:mfg_Law}), infected individuals are self interested and does not undergo mitigation efforts. When the infection is near its peak, susceptible individuals undergo protective measures to avoid getting the infection. This is consistent with the free riding phenomenon which has been observed in the context of COVID-19 pandemic.

\section{Current State-of-the-Art in Optimal Control of Epidemics}
Optimal control of epidemics concerns how a regulating authority can optimally contain epidemic spread in a population. Depending on the underlying epidemic model, we can classify them as (1) compartmental models (2) network based models.
Tools from optimal control were applied to compartmental models as early as 1970s \cite{abakuks1973optimal}, \cite{abakuks1974optimal}.  Compartmental models are formulated as Markov chains and due to their simplicity, they will be the go-to model in our discussions. Nowzari, Preciado and Pappas \cite{NOWZARI16} presents a survey on analysis and control of epidemics in complex networks. 
In this section, we categorize the literature into the following groups according to the control variable available to the regulating authority:
(1) non-pharmaceutical interventions (2) vaccination strategies. 

\subsection{Optimal Control: Non-pharmaceutical Interventions}
The first line of defence against epidemics in the absence of preventative vaccines or in the case of vaccine hesitancy are non-pharmaceutical interventions which reduce the mixing of infected people in the community. Non-pharmaceutical interventions include (1) a full regional lockdown, where only essential services are allowed to operate (2) mandatory use of personal protective equipment
(3) targeted interventions like symptomatic testing, contact tracing and quarantine, and 
(4) travel restrictions.

In the initial stages of the COVID-19 pandemic, the governments all around the world introduced strict regional lockdowns and travel bans to curb the disease spread. However, this led to serious economic and social disruptions as witnessed by the global economic recession following COVID-19.
Early on in the disease evolution, extensive testing, contact tracing and quarantine are effective ways to prevent an exponential growth of the disease. This necessitates a timely intervention from governments in terms of healthcare infrastructure and testing capacity. 
Over the years, researchers have explored how to model the cost associated with these interventions and how to frame the problem faced by the regulator as an optimal control problem.

Abakuks \cite{abakuks1973optimal}, \cite{abakuks1974optimal} first formulated stochastic and deterministic compartmental models where optimal policies are computed numerically for the stochastic model and analytically for the deterministic model.
\cite{abakuks1973optimal} considers a compartmental SIR model of epidemic evolution. The length of infection period is assumed to have an exponential distribution. Under the assumption that a subset of infected population can be isolated instantaneously, the optimal isolation policy is calculated. Note that this is a restrictive assumption because instantaneous isolation is not practical and for many diseases, testing is required to ascertain the state of an individual agent. Optimal isolation policies are numerically computed for a stochastic model and analytically computed for a deterministic model and the two policies compared.
\cite{abakuks1974optimal} uses a similar model to analyze optimal vaccination policies.

Wickwire  \cite{wickwire1975optimal} extended the previous works of Abakuks,  \cite{abakuks1973optimal}, \cite{abakuks1974optimal} on isolation and vaccination to analyze Kermack-McKendrick type compartmental models by means of value function and the Bellman equation.
Optimal isolation policies for deterministic and stochastic epidemics were determined by \cite{wickwire1975optimal}. Wickwire et al. also relaxed the unrealistic assumption that an arbitrary number of individuals can be isolated instantaneously, and instead assumed that there is a hard constraint on the isolation rate.
See \cite{wickwire1977mathematical} for a survey.

Behncke \cite{behncke2000optimal} further adapts Wickwire's model \cite{wickwire1977mathematical} to a more general control and cost setup and deals with both finite and infinite time horizon problems.
They use the Pontryagin's maximum principle \cite{bertsekas2012dynamic} throughout instead of Bellman equations and study the control problem qualitatively, thus avoiding differentiability assumptions on the value function.

In the context of the COVID-19 pandemic, optimal control has been applied to compartmental and network based models. 
Tsay et al. \cite{tsay2020modeling} use a compartmental SEAIR (susceptible-exposed-asymptomatic-infectious-recovered) model with an asymptomatic compartment, provide methods for estimation of parameters, and study the effect of isolation measures.
The work by Perkins et al. \cite{perkins2020optimal} performs optimal control analysis of a compartmental SEAIHV (susceptible-exposed-asymptomatic- infected-hospitalised-vaccinated) model and concludes that (a) heightened control early on in the pandemic is important for achieving long-term success,
(b) preventing a large wave that overwhelms the public health system may not even be possible under some parameter combinations and (c) prioritizing the minimization of deaths versus days under control leads to vastly different outcomes.
Kohler et al. \cite{kohler2021robust} use a compartmental model with 8 states and employ a robust model predictive control (MPC) based feedback policy. This policy adapts the social distancing measures cautiously and safely, thus leading to a minimum number of fatalities even if measurements are inaccurate and the infection rates cannot be precisely specified. 
Silva et al. \cite{silva2021optimal} considers a social opinion biased SAIRP model to provide forecasting mathematical models to anticipate the consequences of political decisions.

Kantner and Koprucki \cite{KANTNER20} compute the optimal non-pharmaceutical intervention strategy based on an extended SEIR (Susceptible-Exposed-Infected-Recovered) model and continuous time optimal control theory. The optimal control must satisfy the the following requirements: (1) minimize disease related deaths; (2) establish a sufficient degree of natural immunity at the end of horizon, to exclude a second wave; and  (3) keep the socio-economic costs of interventions minimum. This model was then calibrated to reproduce the initial exponential growth phase of COVID-19 pandemic in Germany. The optimal intervention strategy can be structured into 3 phases. In the first phase, intervention begins with a strict initial lockdown to hold the effective reproduction number, $\mathcal{R}_{eff} < 1$. In the second phase, there is a \emph{critical period} where the number of simultaneous cases is approximately held constant ($\mathcal{R}_{eff} \approx 1$). During this period, the non-pharmaceutical interventions are relaxed on a gradually increasing rate. Phase 3 commences after the critical period is over and the number of active cases start to decay. In Phase 3,  a final moderate tightening of measures is required. 

Kruse and Strack \cite{KRUSE20} derive optimal policies for social distancing in an SIR model. They find that the optimal policy has the following features: (1) if the death rate is not too sensitive to the number of infected, the optimal policy has two phases. A first phase of strong interventions, followed by a second phase with weaker interventions. (2) If the cost of reducing transmission rate is linear, the optimal policy is always extreme (bang-bang).

Richard and co-authors \cite{RICHARD21} identify optimal age-stratified  non-pharmaceutical interventions to implement as a function of time since the onset of epidemic. By applying optimal control theory, they arrive at a solution which minimizes deaths and control costs. This strategy is implemented for three countries with contrasted age distributions.  They also show that this age-stratified policy strongly outperforms a constant uniform control over the whole population or over the younger population.

Bliman et al. \cite{BLIMAN21} aims to study how partial or total containment can be applied to an SIR epidemic model to minimize the epidemic final size (cumulative number of infected cases during the complete course of an epidemic). Theoretical and numerical results demonstrate that this approach can lead to a significant decrease in epidemic final size. It is shown that optimal intervention has to begin before the number of susceptible individuals has crossed the herd immunity level.

Dimarco, Toscani and Zanella \cite{DIMARCO22} use an SEIAR type compartmental model to study the optimal control strategy when agents reduce their mean number of contacts. The novelty lies in using a kinetic-type model to account for heterogeneity in contact distribution of the population. Using a data driven approach to determine the relevant epidemiological parameters, they show that different types of control can lead to very different mitigation effects according to the level of heterogeneity in contact distribution of agents.

Arruda et al. \cite{ARRUDA21} incorporates multiple viral strains and reinfection to an SEIR model and studies optimal control policies. The model is validated from epidemiology data from COVID-19 in England and Brazil. They consider the cost of mitigation efforts to grow exponentially with the mitigation effort and solve an optimal control problem to to determine optimal mitigation measures. Their results point to the importance of controlling an epidemic from outset in hindering the emergence of new strains and avoiding the effects of a prolonged epidemic.

Morris et al. \cite{MORRIS21} studies the role of non-pharmaceutical interventions in reducing or delaying the peak number of infected individuals. A classical SIR model is used to derive the theoretically optimum strategy and show that easier to implement strategies without perfect information about current state can perform near optimally. However, neither the optimal strategy nor the near-optimal strategies are robust strategies. Small deviations in intervention timing can cause large increases in the infection peak. Robust controls should therefore aim at strong, early interventions which are sustained in an ideal scenario. 

The general conclusion which we can draw from the above  literature is that for linear cost of control, the optimal policy switches from exerting maximum control effort until some point in time and then switches off the control efforts after that time point.
This \emph{bang-bang} solution, with at most one switch, is common in similar problems. Although the bang-bang solution is common, different types of solutions can be obtained for various formulations of the optimal control problem \cite{khouzani2010dispatch}, \cite{khouzani2011optimal}. For an SIR model with quadratic control cost, \cite{khouzani2011optimal} shows that the optimal solution is not a bang-bang controller.


\subsection{Optimal Control with Vaccinations}
Abakuks \cite{abakuks1974optimal} started the work on optimal vaccination strategies for epidemics. However, this model works under the assumption that the whole of susceptible population is instantaneously vaccinated once a vaccine is available. This work computes the optimal control for a stochastic model as well as a deterministic model and provides a comparative study.


Bauch and Earn \cite{BAUCH04} models spread of childhood diseases and analyze voluntary vaccination policies using a compartmental SIR model.
Morton and Wickwire \cite{morton1974optimal} consider an immunisation model for susceptible individuals where the vaccination control is bounded. This boundedness assumption on vaccination control is needed because in practical scenarios, there will be a bound on the maximum number of vaccines available.
Kuga and co-authors \cite{kuga2019vaccinate} combine evolutionary game theory and mathematical epidemiology to evaluate the performance of vaccination subsidizing policies for a seasonal epidemic. Multi-agent simulations are used to find how the topology of network structure affects the vaccination behavior. Mean field approximations are used in this paper to confirm the simulation results and to see the change in social behavior when the vaccine is imperfect. The authors are able to point out instances where vaccine subsidizing policies could be counterproductive. 

Arefin and co-authors \cite{arefin2020mean} build a mean-field vaccination game to analyse the effect of an imperfect vaccine on a two-strain epidemic. The vaccination-decision takes place at the beginning of an epidemic season and depends upon the vaccine-effectiveness along with the cost. An additional situation where the original strain continuously converts to a strain by mutation is also considered.

Zaman, Kang, and Jung \cite{ZAMAN08} propose an SIR epidemic model where a percentage of susceptible population is vaccinated. They show that an optimal control exists for the optimal vaccination problem and describe numerical simulations using a Runge-Kutta fourth order procedure. Furthermore, a real-world example is constructed where smoking is modelled as an epidemic, demonstrating the efficiency of optimal control.
Kar and Batabyal \cite{KAR11} study an optimal control problem with vaccination coverage as a control variable on an SIR epidemic model. With the help of the Pontryagin maximum principle and and an iterative method, it is shown that there are two equilibria, one a disease free equilibrium and the other an endemic equilibrium. The existence and stability of these equilibria are studied and the optimality system is then solved numerically. 

Tchuenche and co-authors \cite{tchuenche2011optimal} analyze the dynamics of an influenza pandemic model with vaccination and treatment using two preventive scenarios: increase in vaccine uptake and decrease in vaccine uptake. The optimal control is computed using Potryagin's maximum principle and  sensitivity analysis and simulations are performed to determine the relative importance of transmission parameters.

Acuna et al. \cite{ACUNA21} formulate an optimal control problem where vaccination coverage (covering a certain percentage of the population in a given period) and hospital occupancy are constrained and identifies vaccination policies that minimizes the number of disability-adjusted years of life lost. A compartmental model is used in the analysis and the burden of COVID-19 is studied with respect to different scenarios such as optimal vs constant vaccination policies, vaccine efficacy, induced vaccine immunity and natural immunity.

\subsubsection*{Note}
It should be noted that although this section focused on optimal control in the context of epidemics, the same models and tools are directly applied to general spreading processes in complex networks. 
Examples of optimal control applied to these models include: malware propagation in computer networks \cite{khouzani2010dispatch}, \cite{khouzani2009optimal}, opinion dynamics in social networks \cite{albi2015optimal}, and adoption of a new product in a marketplace. Nowzari et al. \cite{NOWZARI16} provides a survey on analysis and control of epidemics in complex networks.

\section{Current State-of-the-Art in Mean Field Game Modeling of Epidemics}

This section is devoted to the use of the mean field game approach in the study of epidemic spread and control. We have categorized the relevant literature into the following groups: (1) Analysis of epidemic spread and non-pharmaceutical interventions (2) Study of vaccination effects (3) Control and policy design for Epidemics.

The paper by Huang and Zhu \cite{HUANG22} is a recent survey of the use of game theoretic models (including mean field games) for epidemic spread and control. The paper starts with a review of various models (such as the SIR model) for epidemic spread. The focus of the survey is on the use of the models in answering important questions such as what interventions, when to intervene, etc. The review also provides a taxonomy of the literature based on  (1) types of games, such as static games, differential games, stochastic games, evolutionary games, and  mean field games (2) types of interventions, such as social distancing, vaccination, quarantine, and antidotes; and (3) types of decision-makers, such as individual nodes, adversaries, and central authorities.

\subsection{Mean Field Game Modeling and Analysis of Epidemic Spread and Non-pharmaceutical Interventions}


Elie, Hubert, and Turinici \cite{ELIE20} formulate a model of COVID-19 spread and control using an SIR model with an embedded mean field game. The control in the SIR model is induced by the degree of contact among the individual nodes. An individual node can decrease the contact rate during the epidemic and this intervention has a social cost and an effort cost. The mean field game model is shown to have an equilibrium in which the transmission rate of the epidemic is reduced. The transmission rate achieved in the equilibrium is however higher compared to a socially optimal solution. Similar results are derived in the case of an SEIR model where an additional \emph{exposed} state is introduced. The divergence between autonomous behavior and socially optimal behaviors are shown to be  more prominent immediately before and immediately after the peak of the epidemic.


The paper by Petrakova and Krivorotko \cite{PETRAKOVA22} presents a model for spread of an epidemic like COVID-19 by considering three separate groups of population,  namely, suspectable (S), infected (I), removed (R) and cross-immune (C) ones. The model is based on the mean-field control inside these three groups of population. This model takes into account  population heterogeneity and is therefore superior to a traditional SIR model.  The numerical experiments are able to produce accurate estimates for COVID-19 spread in Novosibirsk, Russia for two 100-day periods.


Cho \cite{CHO20} presents a mean field game model of individual nodes in a population affected by an epidemic, where each node chooses a dynamic strategy of interactions, given the benefits of the interactions as well as the risk involved in getting infected.  The  mean field equilibrium that results from the non-cooperative game model is computed  and the outcome is compared to the socially optimal outcome which maximizes the total utility of the population.  It is shown that the mean field equilibrium strategy is to make more contacts than the level at which it would be socially optimal, in the absence of any public policy or incentives.  If incentives are offered and the cost of  incentivizing people is included, then it is shown that  policies reducing contacts of the infected should continued to be enforced even after the peak of  epidemic has passed. The paper also computes the price of anarchy to  get an idea of  the conditions under which the discrepancies between the mean field behaviour and socially optimal behaviour warrant public policy interventions.


Olmez, Aggarwal, Kim, Miehling, Basar, West, and Mehta \cite{OLMEZ21} develop mean field game models for the evolution of epidemics. The specific problem that is modelled as a mean field game is the decision facing an individual node regarding the degree of social activity. This is modelled as a mean field game, taking into accounts healthcare related cost and benefits accruing from social interactions. The authors investigated the fully observed setting as well as a partially observed setting. The paper presents a complete analysis of the fully observed case and some analytical results for the partially observed case. 
In the fully observed case,  each individual node knows its epidemiological status perfectly. It turns out that a susceptible node will engage in a social interaction if any only the reward outweighs the risk whereas an infected individual will choose to quarantine. In the partially observed case, the nodes do not know their epidemiological status perfectly and it turns out that an infected node behaves like a susceptible node. This could make the epidemic spread faster.
The paper \cite{OLMEZ22} is a follow-up of the  work  in \cite{OLMEZ21} and studies the behavior of self-interested agents in a large heterogeneous population and how pre-symptomatic agents can drive the growth in infection. A mean field type optimal control model is used to investigate the effect of partial observation on individual decision making.


Tembine \cite{TEMBINE20} considers  a class of mean-field-type games with discrete-continuous state spaces and presents Bellman systems that provide sufficiency conditions for  the existence of mean-field-type equilibria in state-and-mean-field-type feedback form. The author derives unnormalized master adjoint systems (MASS) which provide a methodology powerful enough to model the propagation of the COVID-19 virus in the globe. Based on MASS, the author presents data-driven modelling and analytics for mitigating COVID-19. The model is very versatile and captures many aspects:  untested cases, age-structure, decision-making, gender, pre-existing health conditions, location, testing capacity, hospital capacity, and a mobility map of local areas, including in-cities, inter-cities, and international aspects. The author shows that this data-driven model can capture the trends of the reported data on COVID-19.


The report by Bremaud \cite{BREMAUD21} models the propagation of epidemics  in which individual nodes  have control on some parameters such as vaccination rate or social interactions.  The report focuses on two models based on the standard SIR model. The first model focuses on vaccination control. The existence of a unique mean field equilibrium is shown and a numerical method is presented for computing the equilibrium. The second model captures the effect of social contacts. Here again, a mean field equilibrium is computed numerically.

Bremaud and Ullmo \cite{BREMAUD22} consider a SIR compartmental model with social structure, where individuals are grouped by age and interact in different settings where individuals have contact with the other (schools, households, community, etc.). The mean field Nash equilibrium is computed in this setting and is compared with the social optimum. They also investigate how an approximation of socially optimum policy can be obtained with social policies like lockdown.


Aurell, Carmona, Dayanikli, and Lauriere \cite{AURELL21a} discuss the modeling of epidemics using graphon games.  A crucial assumption in mean field game theory is that nodes are indistinguishable and interact identically regardless of with whom they interact. In the context of modeling of epidemics, there is often a need to model the diversity of individual nodes  and the variation of their interactions (for example,  travel restrictions, multiple age groups with distinct social behavior and risk profiles, and a wide range of co-morbid conditions, etc.).  Games with a large number of non-identical players can be analyzed with so-called graphon games whenever the network specifying the interactions is dense. A graphon can be viewed as the limit of a dense random graph. The authors develop a framework for epidemic modeling using graphon games and analyze their Nash equilibria.
They provide a sufficient condition for the existence of a Nash equilibrium and propose a numerical approach based on machine learning tools to compute the equilibrium. The paper also presents numerical results on several applications of  compartmental models for epidemics.


In the paper by Kordonis, Logos, and Papavassilopoulos \cite{KORDONIS22}, the authors study a dynamic game model that captures social distancing behaviors during an epidemic, assuming a continuum of players and infection dynamics caused by individual choice. The authors assume a slight variant of the SIR model. The players have incomplete information about their  infection state, and their choice of actions is determined by the individual beliefs on the probabilities of being susceptible, infected, or recovered. The cost of each player is determined by the infection and the contact factor.  The authors show that a Nash equilibrium is guaranteed to exist and develop an efficient computational procedure for the same.  Even when the players have the same parameters, they could exhibit different behaviors. The work studies the effect of various parameters like the vulnerability (co-morbidity for example) of players, the time horizon, and the various interventions on the optimal policies and the costs of the nodes.


The paper by Gao, Li, Pan, and Poor \cite{GAO21} is concerned with accurate modeling of COVID-19 evolution with  mean field evolutionary dynamics (MFEDs) by invoking optimal transport theory and mean field games on graphs.  The authors compute the payoff functions for different individual states from the commonly used replicator dynamics (RDs) and employ them to govern the evolution of epidemics. The authors compare epidemic modeling based on MFEDs with that based on RDs through numerical experiments.  The efficacy of MFEDs is demonstrated by fitting the model to the  COVID-19 statistics of Wuhan, China.  The authors also analyze the effects of one-time social distancing as well as the seasonality of COVID-
19 through the post-pandemic period.


\subsection{Mean Field Game Modeling of the Effect of Vaccinations}
Doncel, Gast, and Gaujal \cite{DONCEL20, GAUJAL21} present a mean field game model under the SIR model when the individual nodes in the population choose when to get vaccinated. The authors prove the existence of a unique mean field equilibrium which shows a bang-bang control behavior. Specifically, there is a threshold time until which the individuals will get vaccinated at maximal rate and beyond which they do not get vaccinated at all. Interestingly,  the vaccination strategy that minimizes the total cost has the same structure as the mean field equilibrium.  However, the vaccination period of the mean field equilibrium is always smaller than the one that minimizes the total cost.  This essentially means that vaccination should be subsidized appropriately in order to nudge the people to exhibit optimal vaccination behavior.


Hubert and Turinici \cite{HUBERT18} consider an SIR model with vaccination where the vaccination is not mandatory. In particular, they study newborn vaccination. The evolution of each individual node is modelled as a Markov chain and the dynamics of the population is  modelled using a mean field approach. The vaccination decision optimizes a criterion depending on the time-dependent societal vaccination rate and the future epidemic dynamics. It is shown that a Nash mean field equilibrium exists in the proposed model. The paper presents a method for numerical computation of the equilibrium.


Laguzet, Turinici, and Yahiaoui \cite{LAGUZET16} analyzes individual vaccination strategies for a Susceptible-Infected-Recovered (SIR) model. The decision on whether or not to get vaccinated is taken by individual nodes and is based on the risk of infection, the possible side effects of the vaccine and the overall severity of the epidemic course. An important realistic condition is captured, namely limited capacity of vaccination. It is shown that a mean field equilibrium exists between the individual decisions and the epidemic evolution. This enables one to compute an optimal vaccination policy.


Salvarani and Turinici \cite{SALVARANI18} model the behavior of individual nodes in protecting themselves against an epidemic when the vaccination is voluntary. They capture two aspects of vaccines namely imperfect efficacy and limited duration of protection. They show the existence of a Nash equilibrium, assuming non-persistent immunity. They also provide a numerical method for computing the equilibrium. Numerical experiments reveal interesting insights which are useful in planning a vaccination campaign (timing, etc.) in the realistic scenario of imperfect vaccines.


Lee, Liu, Li, and Osher \cite{LEE22} deals with the problem of transporting and distributing the COVID-19 vaccine(s),  to achieve optimal control of the pandemic.  The methodology is based on  a mean-field variational problem in a spatial domain, which controls the propagation of pandemic by the optimal transportation strategy of vaccine distribution.  The authors integrate the vaccine distribution into the mean-field SIR model designed by the authors in \cite{LEE21}. Experimental results show that the proposed model indeed provides effective strategies in vaccine distribution on a spatial domain.

\subsection{Mean Field Game Approach to Control and Policy Design for Epidemics}

Aurell, Carmona, Dayanikli, and Lauriere \cite{AURELL22} model the spread and regulation  of an epidemic as a Stackelberg game between the federal Government and the population. The Government which formulates mitigation policies through incentives is the leader in the game while the mean  field of nodes (citizens) representing the population is the follower. A standard SIR model is considered to represent the spread of the epidemic.  The paper considers for the first time, a compartmental model of epidemics capturing the interplay between independent autonomous nodes and a regulator. The individual nodes interact via a noncooperative game where utility maximization is equivalent to minimizing the individual cost by controlling the  degree of interactions with other nodes (rates of transitions between the states).  The Nash equilibrium of this non-cooperative game is determined. The leader, guided by a social objective, applies incentive policies and non-pharmaceutical interventions which determine the Nash equilibrium of the mean field game among the individual nodes.

The mean field game proposed is an extended MFG in which dependency on the joint distribution of action and state is explicitly taken into account (in earlier models of extended MFG, this dependency involved only the marginal distributions). In terms of the technical novelty of the model,  this paper formulates the leader’s problem under the constraint of the mean field Nash equilibrium of the individual nodes as an optimal control problem with two forward stochastic equations.  The  numerical solution of this problem uses an approximation of the population by an interacting particle system and the approximation of controls by neural networks, including the leader’s policy. The optimization of the leader’s cost is carried out using a variant of stochastic gradient descent to update the neural networks’ parameters (see \cite{AURELL22} for details).

The numerical experiments reveal two interesting insights: (a) In the first case, when the leader applies a containment policy, it is found that the nodes are more cautious about their interactions in the resulting Nash equilibrium when compared to a free spread scenario without any containment policy in place. An early lockdown has a higher impact than any action taken at a later point in time. (b) When the leader optimizes over its policies to minimize its own cost,  the Nash equilibrium of the nodes shows interesting outcomes. For example, when the SIR model is expanded to include two additional states namely Exposed (E) and Deceased (D),   the nodes are found to choose lower contact levels  in their Nash equilibrium than  recommended by the regulator.


Hubert, Mastrolia, Possama\"i, and Warin  \cite{HUBERT22} investigate the optimal control
of an epidemic by offering incentives to lockdown and testing. The interaction between the Government and the individual nodes in the population is modelled as a principal–node problem with moral hazard. This results in a Stackelberg game model. The spread of the epidemic is modelled by stochastic model with  SIS or SIR compartments.  The transmission rate of the epidemic is proposed to be controlled by decreasing the contact rate between individuals (through various means such as reducing physical interactions). This causes a social cost as well as monetary cost to the individual nodes. The paper proposes that the government can aid this through a tax or subsidy as well as implement a testing policy. The testing policy will enable to determine accurately the spread of the epidemic, facilitating isolation of  infected individuals.  The work derives an optimal form of the tax which is  indexed on the (a) proportion of infected individuals and (b) the optimal effort of the population, namely the transmission rate chosen in response to this tax.  This yields an optimization problem to be solved by the government, namely solving an appropriate Hamilton–Jacobi–Bellman equation. Experimental results show that the imposition of a tax policy will induce the individual nodes to cut down on their interactions.

The paper also studies the influence a testing policy can have on limiting the spread of the epidemic. If the testing is done aggressively, then individuals who have tested positive can be isolated. This will enable the individual nodes to interact more freely than if there were no testing and no isolation of individuals who tested positive.


Charpentier, Elie, Lauriere, and Tran \cite{CHARPENTIER20} consider an  extended SIR model with several realistic features of COVID-19 pandemic incorporated. The authors derive an optimal policy for controlling the spread of the epidemic  using and taking into account: (a)  lockdown intervention as well as detection and isolation intervention, (b)  the trade-off between the sanitary and the socio-economic cost of the pandemic, and (c) limited capacity of intensive care units (ICUs).  A detailed sensitivity analysis is carried out with parameters chosen from the COVID-19 literature.  It is shown that the optimal lockdown policy is structured into 4 phases:  (1) A quick and strong lockdown intervention to stop the exponential growth of the contagion;  (2) A short transition to reduce the prevalence of the virus; (3) a long period with full ICU capacity and stable virus prevalence; (4) a return to normal social interactions with disappearance of the virus. This optimal scenario avoids the second wave of infection, provided the lockdown is released sufficiently slowly. It is also shown that with aggressive testing followed by isolation of infected individuals, social distancing norms can be relaxed.


Lee, Liu, Tembine, and Osher \cite{LEE21} presents a mean field game model to control the spread of epidemics in the spatial domain. A standard SIR model is considered and the spatial velocities in the three states S,I, and R are chosen as the control variables. The authors consider three crowds,  Susceptible, Infected, and Recovered,  which evolve spatially.  This is due to mobility, interactions, etc. The central planner seeks to mitigate the risk of infection by controlling the spatial velocity of the nodes in the three states. The paper provides efficient algorithms based on proximal primal-dual methods for obtaining the solutions. It is shown that the proposed model can be effectively used for identifying the infected and susceptible populations in a spatial domain.


Xu, Wu, and Topcu \cite{XU2021} discusses three specific COVID-19 epidemic control models: (1) the susceptible, exposed, infectious, recovered (SEIR) model with vaccination control;  (2) the SEIR model with shield immunity control; and  (3) the
susceptible, un-quarantined infected, quarantined infected, confirmed infected (SUQC)
model with quarantine control.  The paper expresses control outcomes using metric temporal logic (MTL) which is a formal specification language. An example of a control outcome  would be: the population immune from the disease should exceed 200 thousand in the next 100 to 200 days \cite{XU2021}. The paper presents methods to synthesize control strategies with MTL specifications and presents simulation results for three scenarios: (a) 
vaccination control for the COVID-19 epidemic; (b)  shield immunity control for the COVID-19; and (c) quarantine control for the COVID-19.

\section{Future Directions}
A prominent direction that has been pursued by many researchers is to investigate how socially optimal control and mean field game based control differ in an epidemic model. The  control policies influence various costs including infection cost, vaccination cost, quarantine cost, etc.
Assuming that these costs can be suitably chosen (with the support of a regulatory authority),  a few researchers have looked into the problem of nudging the individuals of the population in a  way that mean field game control approaches the performance of socially optimal control. Typically, researchers conduct  mean field game analyses and subsequently suggest certain incentives or deterrents based on the associated equilibria. However, the analyses do not take into account how the individuals react to the policies, except for the work by Aurell, Carmona, Dayanikli, and Lauriere \cite{AURELL22} who arrive at an optimal contract by explicitly taking into account the reaction of the individuals through a Stackelberg game type of model.  This is a promising research direction. One specific problem here is to extend the work of \cite{AURELL22} by taking vaccinations into account. 

In the existing literature, the population is treated as consisting of a single group of individuals. In reality, however, there are multiple logical groups such that the weights on the compartmental model would be different for different groups: (a) healthy, robust individuals; (b) individuals with co-morbidities; (c) elderly individuals (which are more vulnerable), etc. It would be good to extend the analysis and control taking these groupings also into account.

The ongoing pandemic is continuously evolving. Research and innovation have led to multiple vaccines and vaccinated people are well protected. However, compliance to vaccination and pandemic appropriate behaviour is not 100 percent. It would be valuable and interesting to capture some of these phenomena in the epidemic models. Data about the ongoing epidemic will be very useful for these studies. The models developed will be much more credible if available data is incorporated appropriately.

Testing plays an important role in controlling the spread of the epidemic and also in better management of the epidemic effects. For infected individuals, early detection is important for medical intervention and prevention of permanent harm to their bodies. It is important to study this problem as a social problem and compute the optimal testing  strategy for the community as a whole through an appropriate model.

Analytical techniques are hard to develop for the problems mentioned in this paper. This offers a formidable technical challenge and will be an interesting research direction. Most existing solutions are numerical. Therefore, another  promising direction will be to develop improved numerical methods.

\bibliographystyle{IEEEannot}
\bibliography{refs}
\end{document}